\documentclass[10pt]{article}
\usepackage{amsmath,amssymb}
\batchmode

\newtheorem{lem}{Lemma}
\newtheorem{thm}{Theorem}

\newcommand{\pr}{\noindent{\bf Proof}. }
\newcommand{\re}{\noindent{\bf Remark}. }

\newcommand{\pa}{\partial}

\newcommand{\hs}{ \hspace{1cm}}
\newcommand{\tr}{\textrm{ tr }}

\newcommand{\al}{\alpha}
\newcommand{\De}{\Delta}
\newcommand{\de}{\delta}

\newcommand{\Ga}{\Gamma}

\newcommand{\La}{\Lambda}

\newcommand{\ep}{\epsilon}
\newcommand{\si}{\sigma}

\newcommand{\cB}{{\cal B}}
\newcommand{\cC}{{\cal C}}

\newcommand{\cO}{{\cal O}}

\newcommand{\cS}{{\cal S}}

\newcommand{\cK}{{\cal K}}

\newcommand{\cM}{{\cal M}}

\newcommand{\cL}{{\cal L}}
\newcommand{\cP}{{\cal P}}

\newcommand{\cZ}{{\cal Z}}

\newcommand{\bbR}{{\mathbb{R}}}
\newcommand{\bbZ}{{\mathbb{Z}}}

\newcommand{\bbS}{{\mathbb{S}}}

\begin{document}

\title{Infinite volume limit   for  the dipole gas   }
\author{ 
J. Dimock\\
Dept. of Mathematics \\
SUNY at Buffalo \\
Buffalo, NY 14260 }
\maketitle

\begin{abstract}
We consider a classical lattice  dipole gas  with low  activity   
and show that the pressure  has    a limit as   the volume goes to infinity.
The result is obtained  by a renormalization  group  analysis of the model.
\end{abstract}

\section{Introduction}

\subsection{overview}
We study a dipole gas   on a unit lattice  $\bbZ^d$ with     $d \geq 3$.
The  potential between  unit  dipoles with moments  $p_1, p_2 \in \bbS^{d-1}$  
at positions  $x,y  \in \bbZ^d$ is  
\begin{equation}
(p_1 \cdot  \pa)   (p_2 \cdot  \pa)  C  (x-y)
\end{equation} 
where   $C(x-y)$ is the Coulomb potential, that is   the kernel of the inverse Laplacian  
\begin{equation}    
C(x,y)   =  (- \De)^{-1}  (x,y) =(2 \pi)^{-d}   \int_{[-\pi,\pi]^d}  \frac{e^{ip(x-y)}}{  2  \sum_{\mu} (1- \cos p_{\mu})} dp
\end{equation}

For this potential we   consider the dipole gas   in the   grand canonical ensemble.
Let      $\La_N  \subset  \bbR^d$ be a box of the form 
\begin{equation}
   \La_N  =  \left[ \frac{-L^N}{2},  \frac{L^N}{2}\right]^d
   \end{equation}
  where   $L$ is large, odd,  and positive. 
  For    $\La_N  \cap  \bbZ^d$  the  grand canonical 
 partition  function  with  activity  $z>0$ and (for convenience)   inverse temperature  $\beta =1$
can be   represented  as a Euclidean field theory   and     is   given  by   
 \begin{equation}   \label{partition1}
Z_N   =  \int    \exp  \Big(zW( \La_N, \phi)  \Big)   d\mu_{ C } (\phi)
\end{equation}
where   
\begin{equation}  \label{four}
W( \La_N, \phi)   =  2  \int_{\bbS^{d-1}}  dp    \sum_{x \in  \La_N \cap  \bbZ^d}      
  \cos  (   p  \cdot  \pa \phi(x)  )
\end{equation} 
Here     $dp$  is  the normalized rotation invariant measure on  $\bbS^{d-1}$.
The fields     $\phi(x)$    are a    family of   Gaussian random  
variables  indexed by  $x \in \bbZ^d$  with mean zero and covariance   given by the 
positive  definite function    $C(x,y)$.  The measure   $\mu_C$  is the underlying measure.
To  make the connection  with the dipole gas one expands the exponential  in  (\ref{partition1})  
 and carries out the Gaussian integrals.
Similarly one  can  define    correlation functions in terms of the field theory.
  
  One  would like to take the thermodynamic  limit for these quantities,  that is the    limit  as  $N \to  \infty$.
Actually  $Z_N$  itself has no limit  but   there should be a limit for the pressure defined      by   
\begin{equation}
p_N   =  | \La_N|  ^{-1}    \log Z_N  
\end{equation}
as well   as for the  correlation functions.  Such limits have been obtained  by   Fr\"{o}hlich and Park   \cite{FrPa78}   and by  Fr\"{o}hlich and Spencer  \cite{FrSp81}
using a method of correlation inequalities.

In   this paper  we  want  to study    the problem  by a  more  robust   method   which   
is  capable   of    answering  other questions  about the long distance behavior of the model   such as  
decay of  correlations.   If   the potential were integrable   one could 
establish   such   results     with a Mayer expansion.   However the  long distance behavior
 $ \pa_{\mu}  \pa_{\nu}  C(x-y)   =    \cO(  |x-y|^{-d})   $
is not integrable. 
Insead we use the  method of the renormalization group  (RG).   The basic idea
is  to break up the integral  into  a  sequence  of  more controllable  integrals   
and  analyze  the effects separately  at each  stage.

 We  follow  particularly  a   RG  approach   for low activity   recently developed   by  
Brydges  and  Slade    \cite{Bry07},     \cite{Bry09}.     A collateral benefit of this paper 
is  to work out some details of their  method  in  case of the dipole gas.  Earlier work  on  the   RG  approach to the dipole gas
 can be found in   
 Gawedski and Kupiainen
  \cite{GaKu84},  
 Brydges and Yau  \cite{BrYa90},         Dimock and Hurd   \cite{DiHu92},  and    Brydges and Keller
    \cite {BrKe93} .      
    
 In  all   these  treatments  the model is  either   defined on     the   torus   $\bbR^d/ L^N\bbZ^d$ with a momentum 
 cutoff or  on  a toroidal lattice  $\bbZ^d/ L^N\bbZ^d$.   One   obtains bounds  on   the partition 
  function and correlation functions    uniform in $N$.   As explained   above   we   work  on  $\bbZ^d$   with the interaction confined to a finite volume  $\La_N$.    We essentially   
reproduce  the basic    torus results, at least for the partition function,    but then  also   take the   $N \to  \infty$   limit.
The  $N \to  \infty$   limit     would be  awkward for a  sequence of  tori because the $N$  dependence   appears in the 
covariance   $C$ as well as the interaction.   Furthermore for the tori   there are difficulties connected with the change    in topology.      The  disadvantage   for us is that      our  finite volume approximation loses some  translation invariance because of the boundary.    Since translation invariance is a key ingredient in the proof,  dealing with  this loss  is one of the main issues.

 Besides the dipole gas papers mentioned  above  we cite some other papers  which  treat infrared problems
 by RG    techniques.   There is   the work of   Brydges, Dimock, and Hurd   \cite{BDH98a},\cite{BDH98b},  Brydges, Mitter, and Scoppola \cite{BMS03},  and  Abdesselam \cite{Abd07}  on non-Gaussian fixed points  
for   $\phi^4$  models,   and    Dimock and Hurd  \cite{DiHu00}  on Sine-Gordon models in  $d=2$
(the Coulomb gas),  and  Mitter and  Scoppola  \cite   {MiSc08} on self-avoiding random walks. 
These papers   either  either work in a finite volume and get
 bounds   uniform in the volume    or else work with a formal infinite volume limit.   
The  hope is that the techniques of the present paper  point the way  to   carrying  these results over   to an actual 
infinite volume  limit.

\subsection{the main result}  

We  now state the main result.   For our renormalization  group approach  we use   a   different 
finite volume  approximation  than   (\ref{partition1})   following the analysis of    Brydges   \cite{Bry07}.    We  first 
add   a term   $(1- \ep)V( \La_N,  \phi  )$ where  
\begin{equation}  \label{sro}
V(\La_N,  \phi  )  =   \frac{ 1}{4} \sum_{  x  \in  \La_N  \cap  \bbZ^d}  \sum_{\pm \mu =1}^d   
 ( \pa_{\mu} \phi(x) )^2 
  \end{equation}
Here  $\pa_{\mu} \phi$  is either the forward or backward   lattice derivative along the unit  basis vector $e_{\mu}$  defined  by 
\begin{equation}
\begin{split} 
 \pa_{\mu} \phi(x)  = & \phi(x + e_{\mu}) -\phi(x )  \\
   \end{split}
  \end{equation}
  where   $e_{-\mu}  = - e_{\mu}$.  Then  $\pa_{\mu}$ and  $\pa_{-\mu}$  
 are adjoint to each other and   $-\De  =  1/2  \sum_{ \mu} \pa_{\mu}^*  \pa_{\mu}$.
 \footnote{We distinguish forward and backward derivatives to facilitate a symmetric 
 decomposition of  $V(\La_N)$ into  blocks}
 
This  addition of   $(1-\ep )V( \La_N,  \phi  ) $  is partially    compensated  by replacing the covariance   $C$ by  $\ep^{-1} C$.
Thus  instead of   (\ref{partition1})  we consider
\begin{equation}   \label{partition 2} 
Z'_N =  \int      \exp  \Big(z W( \La_N,   \phi)-(1-\ep) V(\La_N,  \phi  ) \Big) 
  d\mu_{\ep^{-1} C } (\phi)  
\end{equation}
Then divide by  
\begin{equation}   \label{partition 3} 
Z''_N =  \int      \exp  \Big(-(1- \ep)V( \La_N,  \phi  ) \Big) 
  d\mu_{\ep^{-1} C } (\phi)  
\end{equation}
and form    a new  finite  volume  partition function
\begin{equation}   \label{partition 4} 
Z_N  =   Z'_N /Z''_N 
\end{equation}
Since  formally   $(Z''_N )^{-1}     \exp  \Big(-(1- \ep)V( \La_N  ) \Big) 
  d\mu_{\ep^{-1} C } $  converges to  $d \mu_C$,   so  formally  $Z_N$
yields   the same limit as   (\ref{partition1}).     This holds  for any choice of  $\ep$; the choice 
of $\ep$ is a choice of how  much  $(\pa \phi)^2$  one is putting in the measure and how much
in the interaction.

The point of the adjustment is that  one can make a shrewd choice of  $\ep$  to facilitate
the  analysis.
The main  result  is:

\begin{thm} \label{one}
 For     $|z|$  sufficiently    small  there is a    $\ep  =  \ep (z)$  close  to 1
so  that   the  pressure  
$
p_N  =     | \La_N|  ^{-1}    \log Z_N   
$
has a limit as  $N  \to  \infty$.   \end{thm}

The  proof will involve  a demonstration  that  with the proper   choice of  $\ep  =  \ep(z)$  the density 
 $ \exp \Big(z W-(1-\ep)V\Big) $  tends   to zero under the RG flow leaving  a   measure  like   
  $\mu_{\ep(z)^{-1} C}$  to describe  the long distance behavior of the system.  Accordingly  
  $\ep(z)$   is interpreted  as   a dielectric constant.  To make this remark precise one  would have
  to study the correlation functions  by these methods.   This   seems   quite feasible, but we do not develop 
  this aspect.

For  the proof  of the theorem  it is convenient  to  rewrite the  partition function.
We  first    scale    $\phi \to   \phi/\sqrt{ \ep}$   and then put   
  $\si   =   \ep^{-1} -1$.      
 Then we have  
\begin{equation}   \label{measure3}
\begin{split}
Z'_N(z,\si)= & \int      \exp  \Big( zW(\La_N,   \sqrt {1 +\si } \phi)-   \si V( \La_N, \phi)\Big)   d\mu_{ C } (\phi)  \\
Z''_N(\si)= & \int      \exp  \Big(-  \si  V( \La_N, \phi))\Big)   d\mu_{ C } (\phi)  \\
Z_N(z,\si)=&Z'_N(z,\si)/Z''_N(\si)  \\
\end{split}
\end{equation}
Then    the problem  is  to   show that  for      $|z|$  sufficiently    small  there is a  (smooth)     $\si   =  \si (z)$  near zero     
such that  with  this choice of  $\si$
\begin{equation}  \label{split}
 | \La_N|  ^{-1}    \log Z_N(z, \si(z)) 
=    | \La_N|  ^{-1}    \log Z'_N(z, \si(z))   
- | \La_N|  ^{-1}    \log Z''_N (\si(z)) 
\end{equation}
has   a limit as  $N \to \infty$.   The two terms are treated separately  and    theorem \ref{one} is 
  proved by taking $\ep(z)  =  (1 +  \si(z))^{-1}$.

The  paper is    organized as follows.  In section \ref{norm}  we  show that   the normalizing factor   $ | \La_N|  ^{-1}    \log Z''_N $ has  a limit.   In  section  \ref{prelim}
we  give  some  general  definitions and estimates  and  define the basic  RG
transformation.   In   section \ref{rg}   we  perform  the detailed analysis of the RG   transformation
isolating   the leading terms.   In   section   \ref{flow}    we   study  the flow of the renormalization
group   and   find  the   stable  manifold   $\si  =  \si(z)$.    Finally  in section  \ref{result}   we   assemble  the
results and prove the limit for  $ | \La_N|  ^{-1}    \log Z'_N   $.

\section{The normalizing factor}     \label{norm}

We  consider the infinite volume limit for the normalizing  factor  $ | \La_N|  ^{-1}    \log Z''_N(\si) $.
This is the problem of  the infinite volume limit for a finite volume perturbation in  the field
strength  and may be of more general interest.

First   we    realize the Gaussian process as given by  $\phi  = C^{1/2} Y$  where 
$Y$ has identity covariance.   Then     
\begin{equation}  \label{coy}
Z''_N(\si)=  \int      \exp  \Big(- \frac{\si}{2}   (Y, T_N Y)      \Big)   d\mu_{ I } (Y) 
\end{equation}
where   $T_N$  is the  positive  operator 
\begin{equation}
T_N  =  \frac 12 \sum_{\pm \mu=1}^d  C^{1/2 } \pa_{\mu}^*  1_{\La_N}  \pa_{\mu}   C^{1/2}
\end{equation}  
and  $1_{\La_N}$ is the characteristic function of  $\La_N$.

\begin{lem}
The  operator      $T_N$  on  $\ell^2(\bbZ^d)$   has the properties
\begin{enumerate}
\item   $ \tr   T_N    =   |\La_N|$
\item    $\|T_N \|  \leq  1$.
\end {enumerate}
\end{lem}
\bigskip

\pr  $T_N$  is trace class since  $1_{\La_N}$ is trace class and  $ \pa_{\mu}   C^{1/2}$ is bounded.   Since  $[\pa_{\mu},  C]=0$ we  have  
\begin{equation}
\tr  T_N   =       \frac 12 \sum_{\pm \mu=1}^d \tr    \Big( \pa_{\mu}^* \pa_{\mu}  C  1_{\La_N}   \Big)
=  \tr   1_{\La_N}   =   |\La_N|
\end{equation}
 The bound   $\|T_N\|  \leq   1$  follows from
\begin{equation}
\begin{split}
| ( h, T_N f) | 
 \leq     &\frac12  \sum_{ \mu} | (\pa_{\mu} C^{1/2}  h)(x)|^2  \chi_{\La}(x) |(\pa_{\mu}C^{1/2}  f)(x)|^2  \\
 \leq   & \left( \frac12   \sum_{  \mu} \| \pa_{\mu} C^{1/2}  h\|^2 \right)^{1/2}
   \left(  \frac12  \sum_{ \mu} \| \pa_{\mu} C^{1/2}  f\|^2 \right)^{1/2} \\
 =  & \|h\| \|f\|\\
\end{split}
\end{equation}
\bigskip

\begin{thm}   \label{two}
For real      $\si$  with   $|\si|<1$, 
$ | \La_N|  ^{-1}    \log Z''_N(\si)$  converges as   $N \to   \infty$.  \end{thm}
\bigskip

\pr
Since   $T_N$  is trace class    and   
\begin{equation}
    \|f\|^2 -  \si    ( f,  T_N f)    \geq   ( 1 -  |\si| ) \|f\|^2 >0
    \end{equation}
    the integral   defining    $ Z''_N(\si)$ in   (\ref{coy})  
  exists  and    can be evaluated  as
 \begin{equation}
 Z''_N(\si) =     \det(1+  \si  T_N)^{-1/2} 
\end{equation}
(See for example  \cite{Sim79}).
 Furthermore   since  $|\si|\|T_N\|  \leq    |\si|  < 1 $ we have the 
expansion
\begin{equation}
  Z''_N(\si)=  \exp \left(    \frac12  \sum_{n=1}^{\infty}  \frac{(-\si)^n}{n}  \tr(  T_N^n) \right)
\end{equation}
(See  for example   \cite{ReSi78}). 
Hence     
\begin{equation}  \label{sonny}
 | \La_N|^{-1}   \log  Z''_N(\si)  = 
  \frac12  \sum_{n=1}^{\infty}  \frac{(-\si)^{n}}{n} \frac{ \tr(  T_N^n)  }{| \La_N|} 
\end{equation}
We have   with the trace norm  $\| \cdot  \|_1$ 
\begin{equation}
|\tr(  T_N^n)|  \leq  \|T_N^n\|_1  \leq     \|T_N\|_1\|T_N^{n-1}\|
\leq   \|T_N \|_1
 \leq     | \La_N|
\end{equation}
Hence  the sum is dominated   by   $\sum_n|\si|^n < \infty$.
  We   show   below   that  for each  $n \geq  1$     
  \begin{equation}  \label{conv}
  a_n   =  \lim_{N  \to \infty} \frac{ \tr (  T_N^n)  }{| \La_N|} 
  \end{equation}
  exists.  Then  by the dominated convergence theorem  
  we have the existence of  
  \begin{equation}
 \lim_{N \to \infty}    | \La_N|^{-1}   \log   Z''_N(\si) = 
  \frac12  \sum_{n=1}^{\infty}  \frac{(-\si)^{n}}{n} a_n
  \end{equation}
  \bigskip

Now  consider   the convergence  (\ref{conv}).     We   write   
 \begin{equation}
\begin{split}
 \tr( T_N^n)    =& 2^{-n}  \sum_{\mu_1, \dots,  \mu_n} 
 \tr   (       1_{\La_N}   \Pi_{\mu_1\mu_2}
 \cdots          \  1_{\La_N}  \Pi_{\mu_n\mu_1}   ) \\
  \end{split}
 \end{equation}
where  the sums   are  over  $\pm \mu  = 1,\dots, d$  and  
\begin{equation}
\Pi_{\mu  \nu}   =   \pa_\mu  C    \pa_{\nu}^*
\end{equation}
We   rewrite   this  as
\begin{equation}
 \tr( T_N^n) =    \sum_{ x  \in  \La_N}     a^N_{n}(x)
 \end{equation}
where   
\begin{equation}
 a^N_{n}(x_1)  =  2^{-n}  \sum_{\mu_1, \dots,  \mu_n}   \sum_{x_2, \dots,  x_n  \in  \La_N} 
    \Pi_{\mu_1\mu_2}(x_1-x_2) \Pi_{\mu_2\mu_3}(x_2-x_3)
 \cdots     \Pi_{\mu_n\mu_1}(x_n-x_1)   
\end{equation}
The quantity  $a_n$ is the same expression  without   the restriction to  $\La_N$.
It is independent of  $x_1$ and we  can take  $x_1=0$.   Thus  it is 
\begin{equation}\label{q}
 a_{n}  =  2^{-n}  \sum_{\mu_1, \dots,  \mu_n}   \sum_{x_2, \dots,  x_n} 
    \Pi_{\mu \mu_2}(-x_2) \Pi_{\mu_2\mu_3}(x_2-x_3)
 \cdots     \Pi_{\mu_n\mu}(x_n)   
\end{equation}

To  see that   $a_n$   is finite  we use  (see   lemma  \ref{multilem} to follow)
\begin{equation}
|\Pi_{\mu \nu}(x-y)|  \leq   C  (  1+  |x-y|)^{-d} 
\end{equation}
then  in  (\ref{q})  we  use the estimate    
\footnote{  To prove it divide the  summation  region   into  $|y|  \leq  |x|/2$   and the complement}
\begin{equation}
\sum_y   (  1+  |x-y|)^{-d} ) (  1+  |y|)^{-d+ k \de  }  \leq  C_{k,\de}  (  1+  |x|)^{-d+ (k+1) \de  } 
\end{equation}
valid for     $k\de  < d$.
Applying this successively to  $x_n, x_{n-1},  \dots $   we  are  left  with   
\begin{equation}
\int   (  1+  |x_2|)^{-2d+ (n-1) \de  }  dx_2
\end{equation} 
which is finite if   $(n-1) \de  <d$.  Thus   $a_n$ is   finite.   Similarly  
one shows   that   $ |a^N_{n}(x)|  $ is bounded   uniformly   in  $N$.

Now we write
\begin{equation}
| \La_N|^{-1}  \tr  (  T_N^n  ) 
=    a_n    +    | \La_N|^{-1} \sum_{ x_1  \in  \La_N}  (   a^N_{n}(x_1)-a_n)
\end{equation}
We   show that   the second  term  above  goes to zero  as  $N \to \infty$ 
to  complete  the proof.

First define a slightly smaller  volume
\begin{equation}
\La_N^*   =   \left[ -  \frac{L^N}{2}   +   N,  \frac{ L^N}{2}  -N\right]^d
\end{equation}
The contribution from  $x_1 \notin \La_N^*$   is    bounded by  
\begin{equation}
  | \La_N|^{-1} \sum_{ x_1  \in  \La_N  -  \La^*_N}   | a^N_n(x_1)-a| 
 \leq \  \cO(1)\frac{  |\La_N  -\La_N^*| }{ |\La_N|  }
  \leq \   \cO(N L^{-N} )
\end{equation}
 which goes to zero.

 Now  suppose that  $x_1 \in     \La^*_N$.     Then we have   
 \begin{equation}  \label{25}
 a^N_{n}(x_1)  -  a_n=  2^{-n}  \sum_{\mu_1, \dots,  \mu_n}   \sum_{(x_2, \dots,  x_n)  \in (( \La_N )^{n-1} )^c} 
    \Pi_{\mu_1\mu_2}(x_1-x_2) \Pi_{\mu_2\mu_3}(x_2-x_3)
 \cdots     \Pi_{\mu_n\mu_1}(x_n-x_1)   
\end{equation}
At least one variable   must  be  in   $\La_N^c$,   say   $x_k$.
Furthermore    at least one   pair of  adjacent   variables must satisfy   
$ |x_j- x_{j+1}|  \geq   N/n$.
 Otherwise    $|x_1 -x_k|   \leq   N(k-1)/n < N$   which contradicts  that  $x_1  \in  \La^*_N,  x_k  \in  \La_N^c$.
 Thus we  can make the estimate 
 \begin{equation}
 | \Pi_{\mu_j\mu_{j+1}}(x_j-x_{j+1})|   \leq    C(1+  | x_j-x_{j+1}|)^{-d}
  \leq  C   (1+  N/n)^{-  \ep }     (1+  | x_j-x_{j+1}|)^{-d + \ep }
  \end{equation}
If   $\ep$ is small enough   the  the reduced   decay does  not affect   
convergence in  (\ref{25}).   Thus we have  
\begin{equation}  
| a^N_{n}(x_1)  -  a_n|   \leq   \cO(N^{-\ep})   
\end{equation}
Therefore
\begin{equation}
\begin{split}
  | \La_N|^{-1} \sum_{x_1 \in    \La^*_N}   | a^N_n(x_1)-a| 
 \leq &   \cO(N^{-\ep})  \frac{  |\La_N^*| }{  |\La_N| }     
\leq     \cO(N^{-\ep})  \\
\end{split}
\end{equation}
 which also goes to zero to complete the proof.

\section{Preliminaries}  \label{prelim}

\subsection{multiscale decomposition}  \label{multi}

Renormalization group  methods  are  based on a multiscale decomposition of 
the basic   lattice   covariance.  We  choose  a decomposition into finite range covariances   developed by 
Brydges,  Guadagni, and Mitter  \cite  {BGM03}.   This is an alternative to   block spin averaging   and   has the  advantage of making fluctuation integrals    simpler  and    the fluctuation covariances   smoother.    The smoothness is essential for the method.

The  decomposition has the form   
\begin{equation}
C (x-y)   =  \sum_{j=1}^{\infty}     \Ga_j( x-y )
\end  {equation} 
where   $\Ga_j(x)  $  is  defined on  $ \bbZ^d$,   is positive semi-definite, and satisfies 
 $\Ga_j (x)   = 0$   if  $ |x|  \geq  L^j/2$  for some odd integer  $L \geq  3$.  Furthermore 
there is a   constant  $c_0$  independent of  $L$  such that  
 \begin{equation}
|  \Ga_j(x)|   \leq   c_0  L^{-(d-2)(j-1)} 
  \end{equation}
 for  all  $j,x$.  It   follows that the series converges uniformly.
 Let  
$\pa^{\al}  =  \prod_{\pm \mu =1}^d  \pa_{\mu}^{\al_{\mu}} $
  be a multi-derivative
and let        $|\al|  =  \sum_{\mu}| \al_{\mu} |$.  
 Then there are  constants  $c_{\al}$  independent of  $L$  such  that 
\begin{equation} \label{imp}
| \pa^{\al}  \Ga_j(x)|   \leq   
 c_{\al}  L^{-(d-2+ |\al|)(j-1) } 
  \end{equation}
Then the differentiated series  converges uniformly   to   $ \pa^{\al} C$.

An elementary consequence of this expansion is an estimate on  the decay  of  $C(x-y)$ as  $|x-y| \to \infty$:

\begin{lem}   \label{multilem}
There are   constants  $C_{L,\al}$  such that 
\begin{equation}    
|\pa^{\al}C(x)|  \leq  C_{L,\al} ( 1 + |x|)^{-d+2- |\al|}
\end{equation}
\end{lem}
\bigskip

\pr     First  consider   the case with no derivatives.    For  $|x| \geq L/2 $  choose   
$k \geq  1 $  so  that   $L^k/2   \leq  |x|   \leq  L^{k+1}/2$.  If   $j \leq  k$   then  $\Ga_j(x) =0$ and we have  
  \begin{equation}  \label{singing}
C (x)   =  \sum_{j=k+1}^{\infty}  \Ga_j( x  )
\end  {equation} 
This is estimated by 
\begin{equation}
 \sum_{j=k+1}^{\infty}  c_0  L^{-(d-2)(j-1)}  
\leq  2c_0  L^{-(d-2)k}  
\leq  c_0L  |x|^{-(d-2)} 
\end{equation}
which suffices.
With derivatives  we get the improved decay from  (\ref{imp}).  This completes the proof.
\bigskip

For the renormalization group  we break off pieces  of  $C(x-y)$  one at a time.  Accordingly we
define 
  \begin{equation}
C_k(x-y)   =  \sum_{j=k+1}^{\infty}    \Ga_j( x-y  )
\end  {equation} 
Then   $C = C_0$  and 
\begin{equation}
C_k (x-y) =   C_{k+1}(x-y)  +  \Ga_{k+1}(x-y)
\end{equation}

\subsection{RG    transformation}  \label{2.2}

The  partition function  (\ref{measure3})  can be written  
\begin{equation}
Z'_N(z,\si)=  \int  \cZ_0^N  (\phi)      d\mu_{C_0} (\phi)
\end{equation}
where  
\begin{equation}  \label{picky}
\cZ_0^N  (\phi)   =  \exp  \Big(zW( \La_N,   \sqrt{1+\si} \phi)- \si  V( \La_N, \phi)\Big)   
\end{equation}
The      identity   $C_0   = C_1 +  \Ga_1$
   lets us replace   an  integral over  $\mu_{C_0}$   by  an integral over
$\mu_{\Ga_1}$  and  $\mu_{C_{1}}$
We have   
\begin{equation}
\begin{split}
Z'_N(z,\si)
=&  \int    \cZ^N_{0}(\phi+  \zeta  )  d  \mu_{\Ga_1} (\zeta)   d \mu_{C_{1}}(\phi)\\
=&  \int    \cZ^N_1(\phi)    d \mu_{C_1}(\phi)\\
\end{split}
\end{equation}
We have   defined  a  new  density   by  the fluctuation integral  
\begin{equation}
 \cZ^N_{1}(\phi)    =   ( \mu_{\Ga_1 } *  \cZ^N_{0})(\phi)   \equiv   \int     \cZ^N_{0}(\phi  + \zeta  )  d  \mu_{\Ga_1} (\zeta)
   \end{equation}

Since      $\Ga_1,  C_1$ are only positive  semi-definite   these are degenerate Gaussian measures.   Nevertheless  these  integrals 
are well-defined   and  the above  
 manipulations  are valid.      We  discuss  these issues in appendix  \ref{App}

Continuing in this fashion we   have the representation  for
$j=0,1,2, \dots  $ 
\begin{equation}  \label{vol}
Z'_N(z,\si)=    \int    \cZ^N_{j}(\phi)      d \mu_{C_j}(\phi)
\end{equation}
where  the density   $  \cZ^N_{j}(\phi) $  is  defined   by  
\begin{equation}  
  \cZ^N_{j+1}(\phi)   =   ( \mu_{\Ga_{j+1}}   * \cZ^N_{j})(\phi)  
   =   \int     \cZ^N_{j}(\phi  + \zeta  )  d  \mu_{\Ga_{j+1}} (\zeta) 
\end{equation}
 Our problem is to study the growth of these densities as  $j \to \infty$.

 Note that we have  refrained from scaling after each fluctuation integral which is the usual
 procedure in  the renormalization group.   Thus the volume stays constant  but correlations  weaken  as we proceed.

\subsection{local expansion}

Each density   $\cZ^N_j(\phi)$  will be written in a form  which exhibits its locality properties known as
a polymer representation.    The  localization becomes coarser  as   $j$  gets larger.

For $j=0,1,2, \dots $   we  partition $ \bbZ^d$  into  $j$-blocks   $B$.  These have side  $L^j$  and   are translates   of 
\begin{equation}
B_0 =  \{ x \in \bbZ^d:   |x|  <  1/2 (L^j-1)\} 
\end{equation}
  by  points in the lattice   $L^j\bbZ^d$.  The set  of 
all  $j$-blocks in  $\La = \La_N$ is denoted  $\cB_j(\La)$ or just  $\cB_j$.  A union  of  $j$-blocks $X$ is called a $j$-polymer.
In  particular  $\La$  is  a  $j$-polymer for   $j  \leq  N$.  The set of all $j$-polymers in 
$\La$  is denoted   $\cP_j(\La)$ or just  $\cP_j$.   The connected $j$-polymers are denoted  $\cP_{j,c}$.

The number  of  $j$-blocks in a  $j$-polymer $X$  is denoted   $|X|_j$.    The $j$-polymer $X$
is a small set     if it is connected  and  $|X|_j \leq 2^d$.  The  set  of all small set   polymers   is denoted  $\cS_j(\La)$
or  just  $\cS_j$.
A  $j$-block  $B$  has a small set neighborhood   
\begin{equation}
B^*   =  \cup   \{  Y \in \cS_j:   Y \supset  B \}
\end{equation}
Similarly a  $j$-polymer  $X$   has a small set  neighborhood  $X^*$.

The  density   $   \cZ^N_j(\phi)$   for  $\phi:  \bbZ^d  \to  \bbR$  
will be  written in the    the general form  
\begin{equation}   
\cZ    =  ( I  \circ   K )  ( \La )  \equiv   \sum_{X  \in \cP_j(\La)}    I(\La - X)  K(X)
\end{equation} 
 The  $I(Y)$ is  a 
background functional  which is   explicitly known  and   carries the main contribution to  the density.    
The  $K(X)$  is  called  a polymer  activity and  represents  small corrections to this background.

We  assume     $I(Y)$  has the  
form   
\begin{equation}
I(Y )  =  \prod_{B \in \cB_j:  B \subset Y}   I(B)
\end{equation}
and that      $I(B, \phi)$    depends  on  $\phi$   only    $B^*$.  
 We also  assume 
$K(X)$    factors   over  the  connected components $\cC(X)$  of   $X$, that  is  
\begin{equation}
K(X)   =  \prod_{Y \subset \cC(X)}    K(Y  )
\end{equation}
and   that  $K(X, \phi)$  only depends on $\phi$  in  $X^*$.

All this is quite general.  Special to our model is the fact 
that the background  $I(B)$   has  the form $ I (E, \si ,  B)   =  \exp(-V(E, \si,  B))$ where 
\footnote{Sums over  $\mu$ are understood to range over  $\pm \mu  = 1, \dots, d$, unless otherwise 
specified}  
 \begin{equation}   \label{sigma}
  V(E, \si ,B, \phi )   =      E(B)   
+   \frac {1}{4}  \sum_{x \in B} \sum_{ \mu \nu  }  \si_{\mu \nu} (B) \pa_{\mu}  \phi(x)     \pa_{\nu}  \phi(x)      
\end{equation} 
for some functions   $E, \si_{\mu \nu}:   \cB_j  \to \bbR$.
In  fact we will usually be able to take   $\si_{\mu \nu}(B)  =  \si  \de_{\mu \nu}$  for 
some constant  $\si$ in which case  
\begin{equation}  \label{64}
 V(E, \si ,B, \phi )   =     
     E(B)   +   \frac {\si}{4}  \sum_{x \in B} \sum_{ \mu}  ( \pa_{\mu}  \phi(x) ) ^2    
     \equiv  E(B)  + \si  V(B)
\end{equation}
Also  in our model    we  will  have   
\begin{equation}  \label{special}
\begin{split}
K(X, \phi)= &  K(X, -  \phi)   \\
K(X, \phi)= &  K(X, \phi+c)   \\
\end{split}
\end{equation}
The second holds   for any constant  $c$  and is equivalent   to saying that   $K(X, \phi)$  only depends   on derivatives   $\pa \phi$.

\subsection{norms}

We define  a menagerie   of norms   following  Brydges   \cite{Bry07}.

\subsubsection{}
If   $X$ is a $j$-polymer   we  consider the Banach space   
$\Phi_j(X)  $   of    functions  $\phi :   X  \to   \bbR$ modulo  constants   with the norm   
\begin{equation}
\|  \phi \|_{\Phi_j(X)}  = h_j^{-1}  \max   \left \{    \| \nabla_j \phi \|_{X,\infty},\ \| \nabla_j^2 \phi \|_{X,\infty} \right  \}
\end{equation}
where 
  \begin{equation}
  \begin{split}
   \| \nabla_j \phi \|_{X,\infty}  =& \sup_{x \in X, \mu}  | \nabla_{j,\mu} \phi(x)|  \\
  \nabla_{j, \mu}  =   L^j \pa_{\mu}  & \hs  h_j  =  L^{-(d-2)j/2} h \\
  \end{split}
  \end{equation}  
Note  that  if  $X$ is also  a  $j+1$ polymer then  we  can consider  $ \|  \phi \|_{\Phi_{j+1}(X)} $.
Since   $ h_j^{-1}  =  L^{-(d-2)/2} h_{j+1}^{-1}$  and    $\nabla_j  =  L^{-1}  \nabla_{j+1}$  we have  
 the contractive property 
 \begin{equation}  \label{66}
  \|  \phi \|_{\Phi_{j}(X)}  \leq  L^{-d/2}   \|  \phi \|_{\Phi_{j+1}(X)}  
\end{equation}

 \subsubsection{}
 Now  consider   polymer activities    
$K(X, \phi )$ for   $X \in \cP_j$. 
  We  assume that     $K(X, \phi )$ only depends on  $\phi $ in  $X^*$ 
   and  is a  $\cC^3$  function on  $\Phi_j(X^*) $.  
 \begin{enumerate}
 \item
For  $n=0,1,2,3$ let   $K_n(X, \phi)  $
be the $n^{th}$  derivative with respect to $\phi$.    It  is a multi-linear  functional  on 
$f_i \in \Phi_j(X^*)$     given by  
\begin{equation}
K_n(X, \phi ; f_1, \dots, f_n)  =
   \frac{ \pa^n}  { \pa{ t_1}  \dots  \pa{t_n}}
K(X, \phi  + t_1 f_1  + \dots  +  t_n f_n) \Big|_{t_i =0}
\end{equation}
  We define   
\begin{equation}
\|  K_n(X, \phi) \|_j   = \sup \{\  |K_n(X, \phi ; f_1, \dots  f_n)|:    \|f_j\|_{ \Phi_j(X^*)}  \leq 1\}
\end{equation}
\item
Next  define
\begin{equation}  \label{Taylor}
\|   K(X, \phi)  \| _j    =   \sum_{n=0}^{3}\  \frac{1}{n!}\  \|  K_n(X, \phi)  \|_j  
\end{equation}
This  combination of derivatives has the multiplicative property  
\begin{equation}
\|K(X, \phi) H(Y, \phi) \|_j    \leq  \|K(X, \phi)\|_j   \|H(Y, \phi) \|_j  
\end{equation}
\item  
Next   we  pick  a large field regulator   $G_j(X, \phi', \zeta)$  which depends on $\phi', \zeta$ in  $X^*$.   It
is assumed to have the form   $G_j(X, \phi', \zeta)=  G_j(X, \phi',0)G_j(X,0,  \zeta)$  and satisfy   $G_j(X, \phi', \zeta) \geq  1$ and     $G_j(X,0,0)=1$.   A  polymer activity   $K(X, \phi)$  is regarded  as  a function  $K(X,  \phi' + \zeta)$ of  $\phi', \zeta$  and we define  a norm 
\begin{equation}
\|  K(X) \|_j     =      \sup_{ \phi', \zeta } \|  K_n(X, \phi' + \zeta) \| _j     G_j(X, \phi', \zeta)^{-1}
\end{equation}
Sometimes  we  want to consider the same norm but with   the polymer activity as a function of $\phi'$ only.
In  this case we put  a prime on the norm and define  
\begin{equation}
\begin{split}
\|  K(X) \|'_j     = &     \sup_{ \phi', \zeta   } \|  K_n(X, \phi') \| _j     G_j(X, \phi', \zeta)^{-1} \\
 =   &   \sup_{ \phi' } \|  K_n(X, \phi') \| _j     G_j(X, \phi', 0)^{-1} \\
 \end{split}
\end{equation}
For large field regulators there  are two     choices.    The strong    regulator is    
\begin{equation}
G_{s, j}(X, \phi', \zeta)  =  \prod_{B  \in \cB_j(X)}  \exp \left(  \|  \phi' \|^2_{\Phi_j  (B^*)} +   \|  \zeta \|^2_{\Phi_j  (B^*)} \right)
\end{equation}
The weak regulator is 
\begin{equation}
\begin{split}
G_j(X, \phi',\zeta)  =& \prod_{B  \in \cB_j(X)}  \exp \left(    c_1 h_j^{-2} L^{-dj}  \| \nabla _j \phi' \|^2_{B,2}
+ c_2 h_j^{-2} \| \nabla^2_j \phi' \|^2_{B^*,\infty} \right) \\
\times  & \exp  \left(   c_3  h_j^{-2}  L^{-(d-1)j}   \| \nabla_j \phi' \|^2_{ \pa X,2}    \right)\\
\times  & \prod_{B \in \cB_j(X)}  \exp \left( c_4 h_j^{-2}   \max_{0 \leq  p  \leq  2} \| \nabla^p_j \zeta \|^2_{B^*,\infty} 
\right)   \\
\end{split}
\end{equation}
(Note that     $ h_j^{-2}   L^{-dj}  \| \nabla _j \phi' \|^2_{B,2}  =  h^{-2} \| \pa \phi' \|^2_{B,2}$
actually has no explicit  $j$-dependence.  Nevertheless it is convenient to write it in this fashion.)
The  norm with   strong regulator is 
denoted $\|  K(X) \|_{s,j}   $, and   the norm with the weak regulator 
is denoted  just  $\|  K(X) \|_j$.   We  note also   (\cite{Bry07},  (6.100)) that 
\begin{equation}  \label{ollie}
G_{s,j}(X)  \leq  G_{s,j}(X)^2  \leq  G_j(X)
\end{equation}
and hence 
\begin{equation}  \label{oliver}
\|  K(X) \|_{j}   \leq   \|  K(X) \|_{s,j} 
\end{equation}

\item 
Finally  for the weak norm   we  define    for  $A  \geq  1$   
\begin{equation}  \label{king}
 \|   K  \| _j  =    \sup_{X \in  \cP_{j,c}}   \|   K(X)  \|_j      A^{|X|_j}
\end{equation}
where the supremum is over connected $j$-polymers $X$.  Polymer activities  $K(X, \phi)$
defined   on connected $j$-polymers   $X \subset  \La_N$  with this norm    constitute a Banach space denoted  $\cK_j(\La_N)$.
\end{enumerate}
\bigskip

\subsubsection{}
The norms are  defined to satisfy the  following properties  which hold for suitable choices of 
$c_1,c_2,c_3,c_4$,  $L$ sufficiently large,  and  $h$  sufficiently large depending on $L$. For the 
proofs see  \cite{Bry07}.
\begin{itemize}
\item   If   $\cC(X)$ are the connected components of $X$ then   
\begin{equation}   \label{queen}
\|  K(X) \|_j     \leq    \prod_{Y  \in \cC(X)}   \| K(Y)\|_j  
\end{equation}
\item   If   $X,Y$   are disjoint (but possibly touching)
 \begin{equation}
  \| \left( \prod_{B \subset X}  F(B)\right) K(Y)\| _j    \leq  \prod_{B \subset X}    \|  F(B) \|_{s,j} \|K(Y)\|_j   
\end{equation}
\item     If  
 \begin{equation}  \label{sharp}
  K^\#(X, \phi)  =  \int  K(X,\phi, \zeta )  d \mu_{\Ga_{j+1}}(\zeta)
 \end{equation}  then   
\begin{equation}  \label{key}
\|K^\#(X) \|'_j    
\leq   2^{|X|_j}  \| K(X) \|_j    \leq    (A/2)^{-|X|_j}\|K\|_j
\end{equation} 
\item  
Suppose that  $U$ is  a  $(j+1)$-polymer  and  hence a $j$-polymer.      Then
\begin{equation}  \label{prince}
\| K(U) \|_{j+1}  \leq    \| K(U) \|'_j
\end{equation}
also for the strong norm.
\end{itemize}

\subsection{estimates}

We illustrate the use of these norms with some estimates we  will need.   
We  work in somewhat  more generality than we need   by  introducing   
potentials of the form  
\begin{equation}
 V( s ,B, \phi )   =  \frac 14  \sum_{x \in B} \sum_{\mu \nu} s_{\mu \nu}(x)   \pa_{\mu}  \phi(x)     \pa_{\nu}  \phi(x)       \\
\end{equation}
The  functions $s_{\mu \nu}(x)$ are normed by  
\begin{equation}     \label{tough}
\|  s \|_j  =      \sup_{B \in \cB_j} |B|^{-1}  \|s\|_{1,B}  =  \sup_{B \in \cB_j} L^{-dj} \sum_{\mu \nu}
\sum_{x \in  B}  |s_{\mu \nu} (x)|
\end{equation}
Note that  if    $s_{\mu \nu}(x)  =  \si \de_{\mu \nu}$   then    $V(s, B)  = \si V(B)$ as defined in  (\ref{64})   and the norm is  $\|s \|_j =2d\ \si$.

\begin{lem}  \label{donkey}  {  \   }
\begin{enumerate}
\item   For any  $s_{\mu\nu}(x)$   
 \begin{equation}    \label{first}
\|   V(s, B)  \|'_{s,j} \leq  h^2 \| s \|_j    \hs \|   V( s, B)  \|_{s,j}      \leq       h^2 \| s \|_j 
\end{equation}
\item  The function    $\si  \to   \exp ( - \si V(B))$ is complex
analytic    and    if       $h^2  \si   $  is sufficiently small
\begin{equation}  \label{second}
   \| e^{- \si V(  B)} \|'_{s, j}  \leq  2    \hs     \| e^{- \si  V( B)} \|_{s, j}    \leq   2
\end{equation}
\end{enumerate}
\end{lem}
\bigskip

\pr   Start with the estimate  for  $x \in B$
\begin{equation}  
|  \pa_{\mu}  \phi(x)  |  =  L^{-j} |  \nabla_{j, \mu} \phi(x)| \leq   h_j   L^{-j}   \|  \phi \|_{\Phi_j(B^*)}  
 =  h  L^{-dj/2}  \|   \phi\|_{\Phi_j(B^*)}
\end{equation}
The first  derivative    is   $ [\pa_{\mu}  \phi(x) ]_1(f)  =   \pa_{\mu} f(x) $  and it  satisfies
$| [\pa_{\mu}  \phi(x) ]_1(f)  | \leq   h  L^{-dj/2}  \| f\|_{\Phi_j(B^*)}$. Hence  
\begin{equation}
\| [\pa_{\mu}  \phi(x) ]_1 \|_j   \leq     h  L^{-dj/2} 
\end{equation}
Adding the derivatives    
\begin{equation}    \label{new}
\|  \pa_{\mu}  \phi(x)  \|_j  \leq      h  L^{-dj/2}    \Big( 1 +   \| \phi\|_{\Phi_j(B^*)} \Big)
\end{equation}

Now   we  estimate
  \begin{equation}  \label{seesaw}
\begin{split}
\| V( s ,B, \phi ) \|_j   \leq  &  \frac 14   \sum_{\mu \nu}  \sum_{x \in B}|s_{\mu \nu}(x)|   h^2  L^{-dj}  
  \Big( 1 +   \| \phi\|_{\Phi_j(B^*)} \Big)^2    \\
\leq   &  \frac 12 h^2  \|s\|_j      \Big( 1 +   \| \phi\|^2_{\Phi_j(B^*)} \Big)  \\
\leq  &    \frac12    h^2  \| s \|_j  G_{s,j}(B, \phi, 0)   \\
\end{split}
\end{equation}
which gives   $\|V(s, B)\|'_{s,j}   \leq     \frac12      h^2  \| s \|_j  $.    Similarly  
\begin{equation}
\begin{split}
\|V(s, B, \phi' + \zeta)\|_j   \leq &  \frac12    h^2  \| s \|_j  (  1 +    \|   \phi' + \zeta \|^2_{\Phi_j(B^*)} ) \\
 \leq  &      h^2  \| s \|_j  (  1 +   \|   \phi' \|^2_{\Phi_j(B^*)} + \|\zeta \|^2_{\Phi_j(B^*)} ) \\
  \leq  &      h^2   \| s \|_j   G_{s,j}(B, \phi', \zeta) \\
\end{split}  
\end{equation}
which gives  $\|   V (s, B) \|_{s,j}   \leq         h^2     \| s \|_j $.

 For the exponential estimates  one can compute the derivatives,  estimate, and resum
 (see      \cite{BDH98a} for details).   Using also  (\ref{seesaw})
 yields  
 \begin{equation}  \label{pokey}
\begin{split}
\frac{3^n}{n!}   \|(e^{- \si V(B, \phi)})_n \|_j
\leq   & \exp  \left( \sum_n \frac{3^n}{n!} |\si|   \|V_n(B, \phi) \|_j  \right)     \\
\leq   & \exp  \left( 9|\si| \|V(B,  \phi) \|_j  \right)     \\
\leq   & \exp  \left(  9  d  h^2   |\si|  (  1 +   \|   \phi \|^2_{ \Phi_j(B^*)}  ) \right)\\
=   & \exp\left(  9d  h^2  | \si|  \right)\ G_{s,j}( B, \phi,0) \\
\end{split}
\end{equation}
Now   multiply  by   $3^{-n}$  and   sum over  $n$ to  obtain for  $3/2 \exp  (  9 d h^2  | \si| )  \leq  2$
\begin{equation}
 \|e^{-\si V(B,  \phi)} \|_{s,j}  \leq  2   \ G_{s,j}( B, \phi, 0) 
\end{equation}  
 which implies $ \| e^{-\si V(B)} \|'_{s, j}    \leq   2$.  The bound   $ \| e^{- \si V(B)} \|_{s, j}    \leq   2$  follows similarly.
  This completes the proof.
\bigskip

We  also need an estimate on the initial interaction.   
In this case  $B \in \cB_0$ is  single site $x$    and we consider
\begin{equation}  \label{syrup}
W(u,B, \phi)   =  2  \int_{\bbS^{d-1}}  dp    
  \cos  (   p  \cdot  \pa \phi(x) u )
\end{equation} 

\begin{lem}  \label{burro}
  {  \  }
 \begin{enumerate}
\item     $W(u,B)  $   satisfies 
\begin{equation}  \label{oneone}
\|W(u, B  )   \|_{s,0}      \leq 2   e^{\sqrt{d}hu}
\end{equation}
$ W(u,B)$  is strongly  continuously  differentiable in $u$. 
\item    $ e^{zW(u, B)}$   is complex   analytic in $z$  and  satisfies
 for   $|z|$ is  sufficiently small  (depending on $d,h,u$)
\begin{equation}  \label{twotwo}
\|  e^{zW(u,  B)} \|_{s,0}   \leq     2  
\end{equation}
 $  e^{zW(u, B)}$  is   strongly  continuously  differentiable in  $u$    
\end{enumerate}
\end{lem}
\bigskip

\pr  (1.)
  A calculation using  $ \sum_{\mu} |p_{\mu}|  \leq  \sqrt{d}$  gives  
  $ \| [ \cos  (   p  \cdot  \pa \phi(x) u )]_n  \|_0  \leq  (\sqrt{d}hu)^n$ and so  
  \begin{equation}  \label{newer}
\|  W (u, B, \phi  )  \|_0 \leq  2 \sup_p \|  \cos  (   p  \cdot  \pa \phi(x) u )  \|_0  \leq  2  e^{\sqrt{d}hu} 
\end{equation}
 This     gives  the required $\|W(u, B  )   \|_{s,0}     \leq 2   e^{\sqrt{d}hu}$.

We  first compute the pointwise derivative in  $u$  which is   
 \begin{equation}
  W'(u,B, \phi )   =   - 2  \int_{\bbS^{d-1}}  dp    
  \sin   \Big (   p  \cdot  \pa \phi(x) u \Big) \Big(  p  \cdot  \pa \phi(x)  \Big)
\end{equation}
Then by  (\ref{new}) at  $j=0$   and  (\ref{newer})  with  sine instead of cosine  
\begin{equation}
  \| W'(u,B, \phi) \|_0   \leq    2     e^{\sqrt{d}hu}   h     \Big( 1 +   \| \phi\|_{\Phi_0(B^*)} \Big)    
\end{equation}
and  hence    
 \begin{equation}   \label{www}
  \| W'(u,B) \|_{s,0}  \leq    4h     e^{\sqrt{d}hu}      
\end{equation}
Higher derivatives are treated similarly.   In particular   for the second derivative
   \begin{equation} 
  \| W''(u,B) \|_{s,0}  \leq  8  h^2     e^{\sqrt{d}hu}      
\end{equation}

To  see that  the pointwise    derivative  is also  the strong derivative  we write 
\begin{equation}
   W(u+ \de,B) - W(u,B)  - \de\  W'(u,B)   =     \int_0^{\de}dt    \int_u^{u+t}  W''(s,B)  ds  
\end{equation}
Inserting the bound on   $W''$    the norm of the expression is   $\cO(\de^2)$  which gives the 
result.   The strong continuity of  $W'$ also follows from the bound on  $W''$.
\bigskip

\noindent (2.)
 For the  exponential bound  
 instead   of the norm  $\| \cdot  \|_0$  with  $G_0$   it suffices to use   the  $G=1$  norm     
 \begin{equation}    \label{mmm}
 \| W(B) \|_{00}  =  \sup_{ \phi  }  \|  W(B, \phi) \|_0 =  \sup_{ \phi', \zeta  }  \|  W(B, \phi' + \zeta) \|_0
\end{equation}
This  is a stronger  norm in the sense that    $\|  W(B)   \|_{s,0}  \leq  \| W(B)  \|_{00}$. We still have   
 $\|W(u, B  )   \|_ {00}    \leq 2   e^{\sqrt{d}hu}$  from  (\ref{newer}).
 The  new norm is  
multiplicative    and so   
\begin{equation}  \label{threethree}
\|  e^{zW(u, B)} \|_{00}  \leq  
 \sum_{n=0}^{\infty}  \frac{|z|^n}{n!}   \| W(u, B) \|_{00}^n    
 \leq    \sum_{n=0}^{\infty} \frac{(2|z| e^{\sqrt{d}hu} )^n}{n!}   
 =  \exp  \Big( 2  |z|  e^{\sqrt{d}hu}  \Big)
\end{equation}
This  implies the same result for   $\|  e^{zW(u, B)} \|_{s,0}$.  

The  pointwise derivative in  $u$  is    $(e^{zW(u)})'   = z W'(u) e^{zW(u)}$
and so  
\begin{equation}
\|(e^{zW(u,B)})' \|_{s,0}   \leq  |z|  \|   W'(u,B) \|_{s,0}  \| e^{zW(u,B)}  \|_{00}   
\end{equation}
which we bound  by     (\ref{www})  and  (\ref{threethree}).
There is a similar  bound on the second derivative  which we use as before  to 
show that the pointwise derivative is a strong derivative.

\section{Analysis of the RG  transformation}  \label{rg}

\subsection{}    We  now explain the Brydges-Slade  RG   analysis,  at the same time  noting
the modifications  due to  boundary  effects.

Suppose we    have   $\cZ( \phi)   =    (I  \circ  K)( \La,  \phi)$  with polymers  on scale  $j$.
This is transformed to  
 \begin{equation}  \label{RG}
  \cZ'(\phi')   =   ( \mu_{\Ga_{j+1}}   * \cZ)(\phi')   \equiv   \int     \cZ(\phi'  + \zeta  )  d  \mu_{\Ga_{j+1}} (\zeta) 
\end{equation}
which we seek to write in the form
$\cZ'( \phi')   =    (I ' \circ  K')( \La,  \phi') $
where   the polymers are now on scale  $j+1$.

Further suppose  we  have picked    $I'$  and  we seek  $K'$ so the identity holds.
Our choice of    $I'$  is taken to have     the   form
\begin{equation}
I'(B' , \phi')   =   \prod_{B \in \cB_j, B \subset  B'}  \tilde I  (B, \phi')          \hs     B' \in \cB_{j+1}
\end{equation}
We  define 
\begin{equation}
\de I (B, \phi', \zeta)  =   I(B, \phi' + \zeta)  - \tilde I (B,  \phi')
\end{equation}  
We also define 
  $\tilde  K = K  \circ   \de I $, more precisely
\begin{equation}
\tilde  K(X, \phi', \zeta) =  \sum_{Y \subset X}  K(Y,  \phi' + \zeta)  \de I^{X-Y} ( \phi', \zeta)   
  \end{equation}     
 For connected $X$  we write
\begin{equation}  \label{check}
\tilde  K(X, \phi', \zeta)  =  \sum_{B \subset  X}   J(B, X, \phi')  +   \check K(X, \phi', \zeta)
\end{equation} 
The quantities  $J(B,X)$  will  eventually   be chosen to depend on  $K$  and to  isolate the 
most important part of  $K$  for  cancellation.  For  now  $J(B,X)$  are free but we require 
 $J(B,X) =0$ unless  $X  \in \cS_j,  B \subset X$   and that    $J(B,X, \phi')$  depend on  $\phi'$ only in $B^*$.
Given  $K$ and  $J$  the equation (\ref{check}) defines   $\check K(X)$  for $X$ connected  and for  any   $X \in \cP_j$   we define
\begin{equation}
\check K(X, \phi', \zeta)  =  \prod_{Y \in \cC(X)}   \check K(Y, \phi',\zeta)
\end{equation}

Then   after using the finite range property
 and making some rearrangements   the representation  $\cZ'( \phi)   =    (I ' \circ  K')( \La,  \phi)$  holds  with  
  (Brydges \cite{Bry07},   Proposition 5.1) 
\begin{equation}  \label{basic}
K'(U, \phi')  =   \sum_{X,\chi   \to  U}   J^{\chi}(\phi')   \tilde  I^{U -(X_{\chi}  \cup X)}(\phi') \check K^\#(X, \phi') 
\hs U \in \cP_{j+1}
\end{equation}
Here
$\chi   =  (B_1, X_1, \dots   B_n, X_n) $  
and the  condition  $X,\chi   \to  U$  is that   
$X_1,  \dots  X_n,  X $  be strictly disjoint   and satisfy
$\overline{ (B^*_1  \cup  \cdots  \cup B^*_n  \cup  X)}  = U$.   Furthermore   
\begin{equation}
\begin{split}
 J^{\chi}(\phi')   = &  \prod_{i=1}^n    J(B_i, X_i, \phi')\\
 \tilde   I^{U -(X_{\chi}  \cup X)} (\phi') =&  \prod_{B \in  U -(X_{\chi}  \cup X)} \tilde   I(B, \phi')\\
\end{split}
\end{equation}
where   $X_{\chi}  =  \cup_i  X_i$.  Finally   $\check K^\#(X, \phi') $  is  $\check K(X, \phi', \zeta )$
integrated over $\zeta$ as in (\ref{sharp}).

At this point  we  have  $K'$   as a function  of  $I, \tilde I, J, K$.  
It  vanishes  at the point $(I, \tilde I, J, K) =  (1,1,0,0)$ since for  $U \neq \emptyset$  we cannot
have   both  $\chi = \emptyset$  and  $X  = \emptyset$.
We  are  interested in  the behavior in a neighborhood  of this point.  We have  the norm (\ref{king})
on  $K$  and   we also define
\begin{equation}
\begin{split}
 \| I\|_{s,j}   = & \sup_{B  \in \cB_j}\| I(B)\|_{s,j}  \\
 \|\tilde   I  \|'_{s,j}    = & \sup_{B  \in \cB_j} \|\tilde  I(B)\|'_{s,j}  \\
 \| J \|'_j    = & \sup_{X \in \cS_j, B \subset X}\|  J(X,B)\|'_j \\
 \end{split}
\end{equation}
Then we have  the following uniform smoothness result.

 \begin{thm}  Let  $A$ be sufficiently large.  \label{three}
 \begin{enumerate}
 \item For         $R>0$    there is a $r >0$  such 
 that the following holds for  all  $j$.    If      $ \|I-1\|_{s,j}< r$,       $\|\tilde I-1\|'_{s,j}< r$,  $ \| J \|'_j  < r $  and $ \| K \|_j  < r $ 
  then  $\| K'\|_{j+1}  <   R$.  Furthermore   $K'$  is a smooth function 
 of  $I, \tilde I,  J, K$ on this domain with derivatives bounded uniformly in $j$.
 \item   
 If also  
 \begin{equation}    \label{zero}
\sum_{X \in \cS_j:    X \supset B}  J(B, X)  = 0
\end{equation}
 then the      linearization   of  $K' =  K'(I, \tilde I, J, K)   $   at   $(I, \tilde I, J, K) =  (1,1,0,0)$  is  
\begin{equation}
\sum_{X  \in \cP_{j,c},  \overline{X}  = U}  \left(
K^\#(X)  +  (I^\#(X)-1)1_{X \in \cB_j}  - (\tilde I(X)-1) 1_{X \in \cB_j}  -  \sum_{B \subset X}   J(B,X)\right)
\end{equation} 
where  
\begin{equation}    \label{sharpsharp}
K^\#(X, \phi)  = \int  K(X,  \phi+ \zeta) d \mu_{\Ga_{j+1}} ( \zeta)
\end{equation}
\end{enumerate}
 \end{thm}

\pr   Brydges  \cite{Bry07}, propositions 5.3  and  6.4.   The proof uses the properties  (\ref{queen})-(\ref{prince}).
For the bounds on derivatives one  can establish analyticity and use Cauchy bounds.

  For the linearization the 
   condition  on $J$  insures that there is no contribution from   $J^{\chi}$.  There is no contribution
 from     $  \tilde  I^{U -(X_{\chi}  \cup X)} $    since   $\chi  =  \emptyset, X = \emptyset$  is not 
 allowed.    The   only contribution is   from  $ \check K^\#(X) $
 and it has the form   stated.
  \bigskip

\subsection{}
Now   we  make some further  specializations. 
 First for a smooth   function $f(\phi)$ on   $\phi  \in  \bbR^{\La}$   let   $T_2f$  denote  a second order
Taylor expansion:
\begin{equation}
(T_2f)( \phi)   =  f(0)  + f_1(0; \phi)  + \frac  12  f_2(0; \phi, \phi) 
\end{equation}
With  $K^\#$  defined in  (\ref{sharpsharp})
we now define   for     $X \in \cS_j$, $ X \supset  B$, $ X \neq B$:
\begin{equation}  \label{A}
J(  B,  X) =    \frac{1}{|X|_j} T_2  K^\#(X)
\end{equation}  
and
 $J(B, B) $  so  (\ref{zero}) is satisfied.   Otherwise     $J(B, X)=0$.  
 \bigskip

We   also    specify    as in  (\ref{64})  that   
\begin{equation}   \label{B}
I(B) =I(E, \si,B) = \exp  (- V(E, \si, B))  
\end{equation}
  and     narrow the choice of  $\tilde  I$   by requiring it  to  have the  same   form  
 \begin{equation}  \label{C}
 \tilde  I(B) =I(\tilde  E, \tilde  \si,B) =  \exp  (- V(\tilde E, \tilde  \si, B))    
 \end{equation}
    with     $\tilde E,  \tilde \si$    still to be specified. Note   that  since  $\sum_{B \subset B'} V(B) = V(B')$ we have that
  $ I'(B') =I( E',   \si',B') =  \exp  (- V(E', \si', B))    $  where  
\begin{equation}
E'(B')  =  \sum_{B \subset B'} \tilde  E(B)  \hs    \si' = \tilde \si
\end{equation}

 Now we have a map   $K' =  K'(\tilde   E,\tilde   \si, E, \si, K)$.  
 As a norm on the energy we take
\begin{equation}
\|  E\|_j  =     \sup_{B \in \cB_j} | E(B)| 
\end{equation}
Then the theorem becomes:

  \begin{thm}  Let  $A$ be sufficiently large. 
 \begin{enumerate}
 \item For    $R>0$    there   is a $r >0$  such that  the following holds for all $j$.  
  If   $\|\tilde  E\|_j,  |\tilde   \si| ,  \|E\|_j,  | \si|,  \| K \|_j  < r$
   then  $\| K'\|_{j+1}  < R$.  Furthermore   $K'$  is a smooth function 
 of $\tilde  E, \tilde  \si, E, \si, K$    on this domain with derivatives bounded uniformly in $j$.
 \item   
 The      linearization   of  $K'$  at the origin   has the form
 \begin{equation}
 \cL_1K   +  \cL_2K   +  \cL_3(E, \si, \tilde E,  \tilde  \si,  K)
 \end{equation}
 where   
\begin{equation}
\begin{split}
\cL_1K(U)  = &    \sum_{X \in \cP_{j,c},X \notin \cS_j, \overline X =U}  K^\#(X)   \\
\cL_2K(U)    =&  \sum_{X \in \cS_j, \overline X =U}  (I -T_2)  K ^\#(X) \\
  \cL_3(E, \si, \tilde E,  \tilde  \si,  K)(U)
  =  &  \sum_{\bar B =U}  \left( V (\tilde E, \tilde \si, B) - V^{\#} ( E, \si, B)
    +  \sum_{X \in \cS_j,X \supset  B}  \frac{1}{|X|_j}  T_2 K^\#(X)  \right)\\
\end{split}
\end{equation}

\end{enumerate}
 \end{thm}
  
 \pr     The new map is the composition of the map   $K' = K'(I, \tilde I, J,K)$ of theorem \ref{three}
 with the maps   $I = I(E, \si),  \tilde I  =  I(\tilde E, \tilde \si),   J  = J(K)$.  Thus it suffices to establish 
 uniform bounds and smoothness for the latter.
 
 For  $I= I(E, \si)$   argue as follows.   First  we   note  that   by     (\ref{second})  
  there is a constant  $c$  such that    the function  $\si  \to   \exp( - \si  V(B) )$   is analytic  in   $|\si|  \leq  ch^{-2}$
 and   satisfies    $\|   \exp( - \si  V(B) )  \|_{s,j}  \leq  2$ on that domain.   Now  if  $|\si|  \leq   ch^{-2}/2$  we can write  
 \begin{equation}
e^{- \si V(B)}  -1     =   \frac{1}{2\pi i}  \int_{|z| = ch^{-2}} \frac{\si \ e^{-z V(B)} }{z(z-\si)} dz
\end{equation}
and estimate  
  \begin{equation}
\|e^{- \si V(B)}  -1 \|_{s,j}  \leq  \frac{2|\si|}{ch^{-2}- |\si|}   \leq  4c^{-1}h^2 |\si|
\end{equation}
Hence 
\begin{equation}  \label{I}
\begin{split}
\|  I(E,\si,B) -1  \|_{s, j}   \leq &  | e^{-E(B)} -1|  \| e^{-\si V(B)}\|_{s,j}    
 +     \| e^{-\si V(B)}-1\|_{s,j}   \\
 \leq &  2 \|E\|_j e^{\|E\|_j}   +  4c^{-1}h^2 |\si|  \\
 \end{split}
 \end{equation}
 and the same bound  holds for  $\|  I(E,\si) -1  \|_{s, j}  $.
 Therefore  for any  $\ep>0$  there is  a $\de >0$  such that  if  $\|E\|_j <\de$ and  $|\si| < \de$
 then   $  \|  I(E,\si) -1  \|_{s, j}  < \ep$  for all  $j$.   The uniform bounds on derivatives can be   verified similarly.   
 
 In the same way   we  show  that    $  \| I(\tilde E, \tilde \si)-1\|'_{s,j}  $  can be  made
 uniformly small  by bounds  on    $\| \tilde  E\|_j$ and $|\tilde \si|$ with uniform bounds on derivatives.

For the linear  map  $K \to  J$  we first  estimate   $ \|  T_2  K^\#(X)  \|'_j $.
As in the proof of   lemma  \ref{donkey} we find that for  $n=0,1,2$
\begin{equation}
\frac{1}{n!}  \|  ( T_2  K^\#(X))_n   (\phi')  \|_j  \leq  2  \| K^\# (X)\|_j  G_{s,j} (X, \phi', 0)
\end{equation}
Summing over $n$ we  get a similar bound for   $ \|   T_2  K^\#(X,\phi')  \|_j $.  Then  by (\ref{ollie}) 
 \begin{equation}
 \|   T_2  K^\#(X)  \|'_j  \leq  \cO(1)  \| K^\#(X)\|'_j  
\end{equation}
By   $(\ref{key})$  this is  bounded  by   $ \cO(1)  \| K\|_j$  
Then   for  $X \neq  B$  we have   $
\|  J(X, B) \|'_j  \leq   \|  T_2  K^\#(X)  \|'_j  \leq    \cO(1)   \| K  \|_j
$.
 The  the same bound holds   for  $\|J(B,B)\|'_j$  and  hence   $\|J \|'_j  \leq    \cO(1)   \| K  \|_j$  which
which  suffices.
 \bigskip

The linearization   is a computation.  Indeed   $J(B,X)$ is designed so that  
\begin{equation}
\begin{split}
&\sum_{X \in \cS_j, \overline{X}  = U}  \left(
K^\#(X)  -  \sum_{ B \subset X}   J(B,X)\right)\\
 =&  \sum_{ \overline{B}  = U}  \sum_{X \in \cS_j, X \supset  B } \frac{1}{|X|_j} T_2 K^\#(X)   
 + \sum_{X \in \cS_j, \overline{X}  = U}   (I - T_2) K^\#(X) \\
 \end{split}  
\end{equation}    
which accounts for the presence of these terms.   Also   the linearization of  $ (I^\#(B)-1)$  is 
$-  V^\#(E, \si, B)$,  and so forth.
This completes the proof.
\bigskip

Next  we   make some estimates on the linearization.

\begin{lem}   Let  $A$ be sufficiently large depending on  $L$.  Then 
the  operator  $\cL_1$  is a  contraction with a norm which goes to zero as  $A \to \infty$.
 \end{lem}

\pr
We   estimate by (\ref{key}), (\ref{prince})
\begin{equation}
\begin{split}
\| \cL_1K(U)  \|_{j+1}    \leq &  \| \cL_1K(U)  \|'_j  
\leq    \sum_{X \notin \cS_j, \overline X =U} \|   K^\#(X) \|_j' 
\leq    \sum_{X \notin \cS_j, \overline X =U}\left( A/2\right) ^{-|X|_j}  \|   K \|_j \\
\end{split}
\end{equation}
Multiply    by   $A^{|U|_{j+1}}$  and take the supremum over   $U$.  This 
yields
  \begin{equation}
\| \cL_1K  \|_{j+1}   \leq \left[ 
 \sup_{U}    A^{|U|_{j+1}}   \sum_{X \notin \cS, \overline X =U}\left(  A/2 \right) ^{-|X|_j}  \right] \|K\|_j
\end{equation}
The bracketed expression goes to zero as $A \to \infty$  (Brydges \cite{Bry07},  lemma 6.18).    Thus for $A$ sufficiently large it is 
arbitrarily small.
 The idea is that for  large  polymers  $X$ such that  
$\bar X = U$  the quantity  $|X|_j$ must dominate  $|U|_{j+1}$.   
  \bigskip

\begin{lem}   Let  $L$ be sufficiently large .  Then 
the  operator  $\cL_2$  is a  contraction with a norm which goes to zero as  $L \to \infty$.
 \end{lem}
\bigskip

\pr   This is exactly  Brydges \cite{Bry07},  proposition 6.11,
   but to account for some differences in notation and for   completeness we include some 
details.  Write    $\cL_2K(U) = \sum_{X \in \cS_j, \overline X =U}  R_X(U)$ where $R_X(U)  =  (I -T_2)  K ^\#(X)$.    We have (\cite{Bry07}, (6.40))
\begin{equation}
 \|R_X(U, \phi) \|_{j+1}  \leq  \left(  1 +  \| \phi \|^3_{\Phi_{j+1}(X^*)} \right)
 \|  K_3^\#(X, \phi)\|_{j+1}
 \end{equation}
 and       by    (\ref{66}) 
 \begin{equation}
 \begin{split}
& \|  K_3^\#(X, \phi)\|_{j+1}
\leq   L^{-3d/2}   \|  K_3^\#(X, \phi)\|_{j} \\
&\leq 3!  L^{-3d/2}      \|  K^\#(X, \phi)\|_{j}  \leq  3!   L^{-3d/2}      \|  K^\#(X)\|'_{j}G_j(X,\phi',0)\\
\end{split}
\end{equation}
and   for  $\phi = \phi'+ \zeta$    (\cite{Bry07}, (6.58))
\begin{equation}
 \left(  1 +  \| \phi\|^3_{\Phi_{j+1}(X^*)} \right)
G_j(X, \phi,0)   \leq   \cO(1)   G_{j+1}(\bar X,  \phi',\zeta)
\end{equation}
Combining these   yields 
\begin{equation}
  \|R_X(U, \phi) \|_{j+1}  \leq  \cO( L^{-3d/2}   )     \|  K^\#(X)\|'_{j}  G_{j+1}(\bar X,  \phi',\zeta)
\end{equation}
and hence   using also  (\ref{key})
\begin{equation}
\|R_X(U)\|_{j+1}   \leq  \cO( L^{-3d/2}   )     \|  K^\#(X)\|'_{j} \leq  
  \cO( L^{-3d/2}   )   ( A/2 )  ^{-|X|_j}   \| K\|_{j} 
\end{equation}
Therefore
\begin{equation}
\|\cL_2 K(U) \|_{j+1}  
 \leq    \sum_{X \in \cS_j, \bar X  = U}  \|R_X(U)\|_{j+1}
 \leq    \cO( L^{-3d/2}   )    \sum_{X \in \cS_j, \bar X  = U}  ( A/2 )  ^{-|X|_j}   \| K\|_{j} 
\end{equation}
and so   
\begin{equation}
\|\cL_2 K \|_{j+1}  
 \leq  \cO( L^{-3d/2}   ) \left[ \sup_U A^{|U|_{j+1}}   \sum_{X \in \cS_j, \bar X  = U} ( A/2 )  ^{-|X|_j} \right]  \| K\|_{j} 
\end{equation}
But the bracketed expression  is  $\cO(L^d)$  (\cite{Bry07}, (6.90))
  so  we have   $\|\cL_2 K \|_{j+1}  
 \leq  \cO( L^{-d/2}   )\| K\|_{j} $ to complete the  proof.

\subsection{}
The term  $\cL_3$  needs a more extensive treatment. 
 First  we  localize  the final term in  $\cL_3$  which is      
\begin{equation}
\sum_{\overline{ B} =U}     \sum_{X \in \cS_j,X \supset   B } \frac{1}{|X|_j} \left( K^\#(X,0) 
  +  \frac12  K^\#_2(X,0;  \phi,  \phi)\right)
\end{equation}
In    $ K^\#_2(X,0;  \phi,  \phi)$  pick a point  $z  \in   B$   replace  $\phi(x)$ by  
\begin{equation}
    \phi(z) + \frac12  ( x - z)  \cdot  \pa    \phi(z)\
 \equiv  \  \phi(z) + \frac12  \sum_{\mu} ( x_{\mu} - z_{\mu})  \pa_{\mu} \phi(z)
   \end{equation}
with the thought that the difference is  irrelevant  \footnote{We need  the factor $1/2$ since
the  sum is over  $\pm \mu = 1, \dots, d$.  The convention is  that   $x_{-\mu} = - x_{\mu} $ }.  However   $\phi(z)$   and  $ z \cdot   \pa \phi(z)$
are  constants and    do not contribute.   Thus we   replace  $\phi(x)$  by  
 $\frac12    x \cdot \pa \phi(z)$.   If we also average over    
 $z \in B$    our expression becomes
\begin{equation}  \label{kumquat}
\sum_{\overline{ B} =U}     \sum_{X \supset   B } \frac{1}{|X|_j} \left( K^\#(X,0) 
  +   \frac {1}{8 |B|  }  \sum_{z  \in B}\sum_{\mu \nu} K^\#_2(X,0;  x_{\mu}, x_{\nu} )  \pa_{\mu}\phi(z) \pa_{\nu}\phi(z)\right)
  +\cL_3'K(U) 
\end{equation}
where    $\cL_3'K(U) $  is the error, namely
 \begin{equation}
\begin{split}
\cL_3'K(U)   = & \sum_{\bar B  = U}   \sum_{X  \in \cS_j: X  \supset  B}
\frac{1}{|X|_j}   \sum_{z \in B}  \frac{1}{|B|}     \left( \frac12   K^\#_2(X, 0 ;  \phi, \phi)-  \frac18
 K^\#_2(X, 0 ; x \cdot \pa   \phi(z), x \cdot \pa \phi(z))  \right)\\
 \end{split}
\end{equation}
 Next    we
 define 
\begin{equation}  \label{sunny}
\begin{split}
\beta(B)  =  \beta(K,B) = & -  \sum_{X \in \cS_j,  X \supset B   } \frac{1}{|X|_j}  K^\#(X,0)  \\
\al_{\mu \nu}(B) =  \al_{\mu \nu}(K,B)  =&     
-  \frac12  \frac{1}{|B|}  \sum_{X \in \cS_j,  X \supset B    } \frac{1}{|X|_j} K^\#_2(X,0; x_{\mu},  x_{\nu})\\
\end{split}
\end{equation}
Note that   $\al_{\mu \nu}$ is symmetric and satisfies  $\al_{-\mu \nu} =  - \al_{\mu \nu}$.
We  also let   $\al_{\mu\nu}$ stand for  the function $\al_{\mu\nu}(x)$  which  takes the constant 
value  $\al_{\mu \nu}(B)$ for   $x \in B$.

Now we  write   (\ref{kumquat}) as  
\begin{equation}
\begin{split}
&-\sum_{\overline{B}  = U} \left( \beta(B) +   \frac  14     \sum_{z \in B}  \sum_{\mu \nu} \al_{\mu \nu} (B) \pa_{\mu} \phi(z) \pa_{\nu}  \phi(z) \right)  +\cL_3'K(U) \\
&=  - \sum_{\overline{B}  = U}   V(\beta, \al, B, \phi)  + \cL_3'K(U) \\
\end{split}
\end{equation}
with   $  V(\beta, \al, B, \phi) $   defined as  in  (\ref{sigma}).  
Altogether then we have 
\begin{equation}  \label{oooh}
 \cL_3(E, \si, \tilde  E, \tilde  \si,  K)(U)=
  \sum_{ \overline{B}  = U} 
  \left( V(\tilde  E, \tilde  \si, B)-V^\#(E, \si, B)  -    V( \beta,  \al,  B)  \right)   +  \cL'_3 K (U)
  \end{equation}

\begin{lem}   \label{albeta}
\begin{equation}
\begin{split}
\|\beta \|_j \equiv  \sup_{B \in \cB_j}  | \beta(B)| \leq &  \cO(1) A^{-1} \|K\|_j\\
\|   \al  \|_j \equiv    \sup_{B \in \cB_j}  \sum_{\mu \nu}  | \al_{\mu \nu}(B) |   \leq &  \cO(1)h^{-2} A^{-1} \|K\|_j\\
\end{split}
\end{equation}
\end{lem}
\bigskip

\re  Note that the norm  $ \| \al \|_j$  agrees with  the norm  $\| s \|_j$ in  (\ref{tough}) if   $s_{\mu \nu}(x)
=  \al_{\mu \nu} (B)$ for   $x \in B$.
\bigskip

\pr   By  (\ref{key}) we have  
\begin{equation}  \label{100}
\begin{split}
|  K^\#(X,0)| 
 \leq  &  \| K^\#(X) \|'_j    \leq   (A/2)^{-1}  \|K\|_j \\
\| K_2^\#(X,0)\|_j
\leq &2 \|   K^\#(X) \|'_j  \leq    A^{-1}  \|K\|_j\\
\end{split}
\end{equation}  
Since the number of  small sets containing a block $B$  is bounded by a constant depending only 
on the dimension we have    
\begin{equation}
|\beta(B)|  \leq     \sum_{X \in \cS_j, X  \supset B}  | K^\#(X, 0) |   \leq \cO(1)  A^{-1}  \|K\|_j
\end{equation}
For the bound on $\al$     note that   $\| x_{\mu}\|_{\Phi_j(X^*)} =  h^{-1}L^{dj/2}$.  Then since
$|B|^{-1}  =  L^{-dj}$ 
\begin{equation}
|B|^{-1} |  K_2^\#(X,0;  x_{\mu} , x_{\nu})|  \leq 
h^{-2}\|   K_2^\#(X,0) \|_j
\leq     h^{-2}  A^{-1}  \|K\|_j
\end{equation}
whence  
\begin{equation}
  \sum_{\mu \nu}     |\al_{\mu \nu}(B) |
\leq    \sum_{\mu \nu}   \sum_{X \in \cS_j,X  \supset B}  |B|^{-1}|  K_2^\#(X,0;   x_{\mu},  x_{\nu})|  
\leq    \cO(1)h^{-2} A^{-1}  \|K\|_j
\end{equation}
which  gives the result

\begin{lem}    \label{prime} Let  $L$ be sufficiently large.   Then 
the  operator  $\cL'_3$  is a  contraction with arbitrarily small norm.
 \end{lem}
\bigskip

\pr 
We  have  
 \begin{equation}
\begin{split}
\cL_3'K(U)   =  & \sum_{\bar B  = U}   \sum_{X  \in \cS_j: X  \supset  B}
\frac{1}{|X|_j}   \sum_{z \in B}  \frac{1}{|B|}     \left(  \frac12   K^\#_2(X, 0 ;  \phi- \frac12  x \cdot   \pa \phi(z),  \phi)  \right)
+  \textrm{ similar }\\
\end{split}
\end{equation}
Since   $\pa_{-\mu} \phi(x)  =  - \pa_{\mu} \phi(x- e_{\mu})$ we have   
\begin{equation}
\frac{\pa}{\pa x_{\mu}}   \left(  \phi(x)  - \frac 12  \sum_{\nu} x_{\nu}  \pa_{\nu} \phi(z)\right)  =
\pa_{\mu} \phi(x)  - \frac 12   \pa_{\mu} \phi(z)  -\frac12 \pa_{\mu} \phi(z- e_{\mu})
\end{equation}
The same holds  with  $\pa_{\mu}$ replaced by   $\nabla_{j, \mu}$  and  
then  with    $\textrm{ diam}_j(X^*)  =  L^{-j}\textrm{ diam}(X^*)   $
\begin{equation}
\|  \nabla_j (  \phi- \frac12  x \cdot   \pa \phi(z))  \|_{X^*, \infty}
\leq   \textrm{ diam}_j(X^*)   \|  \nabla^2_j   \phi  \|_{X^*, \infty}
\end{equation}
But    $\textrm{ diam}_j(X^*)   \  \leq  \cO(1)$ since   $X$ is a small set.
Hence
\begin{equation}  \label{ugly}
\begin{split}
\|   \phi-\frac12   x \cdot   \pa \phi(z)  \|_{\Phi_j(X^*)}  
 \leq &\cO(1)   h_j^{-1}  \|  \nabla^2_j   \phi  \|_{X^*, \infty}  \\
  \leq &   \cO(   L^{-d/2-1} )  h_{j+1}^{-1}  \|  \nabla^2_{j+1}   \phi  \|_{X^*, \infty}    \\
   \leq &  \cO(   L^{-d/2-1} ) \|  \phi \|_{\Phi_{j+1} (X^*) } \\
\end{split}
\end{equation}
Now we  estimate 
\begin{equation}
H_X(U,  \phi)  =    K^\#_2(X, 0 ;  \phi- \frac12  x \cdot   \pa \phi(z),  \phi) 
\end{equation}
We  claim that  
\begin{equation}  \label{sugar}
\begin{split}
|(H_X(U))_0 (\phi) |  \leq &\cO( L^{-d-1})\| K_2^\#(X,0)\|_j  \|  \phi  \|^2_{\Phi_{j+1}(U^*)}\\
\|(H_X(U))_1(  \phi)\|_{j+1}  \leq  &  \cO( L^{-d-1}) \| \| K_2^\#(X,0)\|_j  \|  \phi  \|_{\Phi_{j+1}(U^*)}  \\
\| (H_X(U))_2 (\phi)\|_{j+1}  \leq   & \cO( L^{-d-1}) \| K_2^\#(X,0)\|_j   \\
\end{split}
\end{equation}
For example the second bound follows from (\ref{66})  and  (\ref{ugly}) by
\begin{equation}
\begin{split}
&|(H_X(U))_1( \phi;  f) | \\ = &  |  K^\#_2(X,0;  \phi-\frac12   x \cdot   \pa \phi(z) , f) 
 +    K^\#_2(X,0; f-\frac12   x \cdot   \pa f(z) ,  \phi)   |\\
\leq   &  \|  K^\#_2(X,0) \|_j \left(\|\phi- \frac12  x \cdot   \pa \phi(z) \|_{\Phi_j(X^*)}\|   f  \|_{\Phi_j(X^*)}  + 
 \|f-\frac12   x \cdot   \pa f(z) \|_{\Phi_j(X^*)}\|   \phi  \|_{\Phi_j(X^*)}   \right)\\
 \leq   &\cO( L^{-d-1})  \| K_2^\#(X,0)\|_j \|\phi \|_{\Phi_{j+1}(X^*)} \|   f  \|_{\Phi_{j+1}(X^*)} \\
\end{split}
\end{equation}
To complete the bound we  need    $ \|   f  \|_{\Phi_{j+1}(X^*)}   \leq   \|   f  \|_{\Phi_{j+1}(U^*)}$  which holds provided   $X^*  \subset  U^*$.  Here  $X^*$  is an $\cS_j$ neighborhood of  $X \in \cS_j$ and  $U^*$ is 
an  $\cS_{j+1}$  neighborhood of   $U \in \cB_{j+1}$.

  To  see that  $X^* \subset  U^*$  note first
that   $X^*  \cap   U  \neq  \emptyset $  since both contain $B$.   Suppose       $X^* \subset U^*$ is 
false.  Since  points not  in $U^*$  are separated from points in $U$ by at  least $L^{j+1}$ we have
$\textrm{diam} (X^*)  \geq  L^{j+1}$.   On  the other  hand   $\textrm{diam} (X) \leq  \cO(1) L^j$  so  
  $\textrm{diam} (X^*) \leq  \cO(1) L^j$.
This is a contradiction for $L$ sufficiently   large.
 
Combining these estimates  (\ref{sugar})  we get  
\begin{equation}
\|H_X(U, \phi)\|_{j+1}    \leq     \cO( L^{-d-1}) \| K_2^\#(X,0)\|_j (1 + \| \phi\|_{\Phi_{j+1}(U^*)}^2)  \\
 \end{equation}
 But  for  $\phi  = \phi' + \zeta$
 \begin{equation}
 (1 + \| \phi\|_{\Phi_{j+1}(U^*)}^2)   \leq  G_{s,j+1}(U, \phi,0)   \leq  G_{s,j+1}(U, \phi',\zeta)  
 \leq    G_{j+1}(U, \phi',\zeta)  
\end{equation}
  Using also    (\ref{100})   we obtain
 \begin{equation}
\| H_X(U)  \|_{j+1}    \leq    \cO( L^{-d-1})  A^{-1} \|K \|_j
 \end{equation}
which implies
\begin{equation}
\| \cL'_3K(U)  \|_{j+1}  \leq    \cO(1)   \sum_{\bar B  = U}  \| H_X(U)  \|_{j+1}  
 \leq     \cO( L^{-1})  A^{-1} \|K \|_j
\end{equation}
Since   $ \cL'_3K(U)$ is zero  unless   $|U|_{j+1}  =1$  this gives  
$\| \cL'_3K  \|_{j+1}  \leq       \cO( L^{-1})   \|K \|_j$ which completes the proof.
\bigskip

\subsection{  }

Now  consider the first  term  in  (\ref{oooh}).   We  would like to  choose  $\tilde E, \tilde \si$
so it vanishes  but  are not quite there  yet.

To  proceed we add another hypothesis.   We  assume that  $E(B), K(X, \phi)$   are
invariant under lattice symmetries for  $B,X$  away from  the boundary of  $\La_N$, that is if     
 $B,X$ have no boundary blocks.     More precisely    $E(B)$  is independent of $B$,  and if  $g$ is a translation,  rotation by a multiple of $\pi/2$,    or  a reflection and  $(g\phi)(x)  = \phi(g^{-1} x)$  then  $K(gX, g\phi)  =  K(X, \phi)$   provided 
 $X, gX$ are away from the boundary.

These   properties carry over  to the next level  and  to   the   quantities  $\beta(B),  \al_{\mu \nu}(B)$.

\begin{lem}
Suppose   $E(B),  K(X, \phi)$   are invariant under lattice symmetries   away from the boundary of  $\La_N$
and   $\tilde E(B)$  is  invariant for  $B^*$  away from the boundary. Then 
 \begin{enumerate}
\item   
$ E'(B'),  K'(U, \phi)$  are invariant for  $B', U$   away  from the boundary
\item     If   $B^*$  is away from the boundary then  
$\beta(B),\al_{\mu \nu}(B) $   are      independent of  $B$   and   $\al_{\mu \nu}(B)  =
\hat   \al_{\mu \nu}(B)$
defined for all  $B$ by 
 \begin{equation}     \label{verysunny}
\hat  \al_{\mu \nu}(B)   = \frac{\al}{2}  \   (  \de_{\mu \nu}  -  \de_{\mu, -\nu} )
\end{equation}
where   $\al$ is a constant.
\end{enumerate}
\end{lem}
\bigskip

\pr   If   $B' \in \cB_{j+1}$ is  separated from the boundary  then  $d(B', \pa \La_N)  \geq  L^{j+1}$.  If    $B  \subset  B'$ then  $d(B^*, \pa \La_N) \geq  L^{j+1} -  2^d  \geq  L^j$  so   $B^*$  is away  from the boundary.
Thus in   $E'(B')  = \sum_{B \subset  B'}
\tilde  E(B)$  each  $\tilde E(B)$ is invariant and hence  so  is  $E'(B')$.

Under our hypotheses   $\tilde K(X)$ defined with  (\ref{B}),  (\ref{C}) is invariant for   $X^*$  away from the boundary, 
  and using the invariance of  $\Ga_j$  the quantity   $J(B,X)$
 defined  by   (\ref{A})  is invariant for  $B^*$  away from the boundary.   Thus  $\check K(X)$  is invariant  for   $X^*$  
away from the boundary  and    so  is  $\check K^\#(X)$.  
Now in the definition    (\ref{basic}) of    $K'(U)$   the quantity   $\check K^\#(X)$  only  contributes
for  $X \subset U$.      Then     U  away from the boundary implies     $X^*$   away from the boundary,   so only invariant terms  $\check K^\#(X)$  contribute.  Similarly  only
invariant  terms contribute  to   $J^{\chi}$  and    $\tilde  I^{U -(X_{\chi}  \cup X)}$.   Hence 
$K'(U)$ is invariant.

The quantities  $\beta(B),  \al_{\mu \nu}(B)$  depend on  $K(X)$  for  $X \subset B^*$ so if  $B^*$ is away  
  from the boundary  they are invariant and in particular independent of $B$.   Furthermore  under the same condition if  
 $R$ is a rotation or a reflection we have  for  $\mu, \nu >0$
\begin{equation}    \label{reflect}
\al_{\mu \nu}(  B)   =    \sum_{\mu ' \nu'>0}  R_{\mu \mu'}\  R_{ \nu \nu'}\ \al_{\mu' \nu'}(  B)  
\end{equation}

To establish the identity    $\al_{\mu \nu}(B)  =
\hat   \al_{\mu \nu}(B)$
note that  since  both   are symmetric and satisfy    
 $\al_{-\mu \nu}(B)=  -  \al_{\mu \nu}(B)$  it  suffices to establish the identity for  $\mu, \nu >0$
 in which case it says  $\al_{\mu \nu}(B)  =  \al  \de_{\mu \nu}/2$.  Specializing (\ref{reflect})  to reflections through planes  $x_{\mu} =0$     we deduce that   $\al_{\mu \nu}(B)$  equals  zero unless
$\mu = \nu$  so   $\al_{\mu \nu}(B)  =  \al_{\mu} \de_{\mu \nu}/2$.    Specializing (\ref{reflect})  to rotations 
 we  deduce  that    $\al_{\mu}$ is independent of  $\mu$ and obtain the  result.   This completes the proof.
\bigskip

  We also define  for all  $B \in \cB_j$
\begin{equation}
\al'_{\mu \nu}(B)   =   \al  \   \de_{\mu \nu}
\end{equation}
and    write    for any  $U \in \cB_{j+1}$
 \begin{equation}
 \sum_{\overline{B} =U} V( \beta,  \al,  B)   =   \sum_{\overline{B} =U} V( \beta,  \al',  B)    -  \cL_4K(U)   -   \De K(U)   
 \end{equation}
 where   for  $U \subset  \cB_{j+1}$
 \begin{equation}
 \begin{split}
  \cL_4K (U)  = & \sum_{\overline{B}=U}  V(0,  \al' -  \hat \al    ,  B)  =   V(0, \al'  -  \hat \al  , U)  \\
 \De K (U)  = & \sum_{\overline{B}=U}  V(0, \hat \al  - \al  ,  B) = V(0,  \tilde  \al,  U)        \\
 \end{split}
 \end{equation}
 where  $\tilde \al_{\mu \nu}  (x)   =   \hat   \al_{\mu \nu}  (B)- \al_{\mu \nu}  (B) $  if  $x \in B$. 
Note that  $\De K(U)$    vanishes  unless  $U$ touches the boundary.     Now     (\ref{oooh})  becomes
\begin{equation}   \label{grrr}
\begin{split}
& \cL_3(E, \si, \tilde  E, \tilde  \si,  K)(U) \\
=&
  \sum_{ \overline{B}  = U} 
  \left( V(\tilde  E, \tilde  \si, B)-V^\#(E, \si, B)  -    V( \beta,   \al',  B)  \right)   +  \cL'_3K (U)
  +\cL_4K(U)  +  \De K  (U)\\
  \end{split}
  \end{equation}
\bigskip

\begin{lem}   Let  $L$ be sufficiently large.   Then 
the  operator  $\cL_4$  is a  contraction with arbitrarily small norm.
 \end{lem}
\bigskip

\pr  
For  $U \in \cB_{j+1}$  
 \begin{equation}
  \cL_4K(U)   =    \frac{\al}{8}      \sum_{\mu}  \sum_{x \in  U}    
       \pa_{\mu} \phi(x)\  \pa_{-\mu} \phi(x) +  \pa_{\mu} \phi(x) ^2  
\end{equation}
But   $\pa_{- \mu}  \phi(x)  = -  \pa_{\mu} \phi(x- e_{\mu})$     and  $   \pa_{\mu} \phi(x- e_{\mu})-   \pa_{\mu} \phi(x)  =  
-(\pa_{- \mu}  \pa_{\mu} \phi)(x)$
so this is  
\begin{equation}
  \cL_4K(U)   =    -   \frac{\al}{8}     \sum_{\mu}  \sum_{x \in  U}    
     (\pa_{- \mu}  \pa_{\mu} \phi)(x)    \pa_{\mu} \phi(x)  
\end{equation}
The proof now proceeds as in  lemma  \ref{donkey}  but now on scale  $j+1$.   Instead of  (\ref{new})
we have   
\begin{equation}
\begin{split}
|  \pa_{\mu}  \phi(x)  |  \leq &  h  L^{-d(j+1)/2}  \|   \phi\|_{\Phi_{j+1}(U^*)}\\
|\pa_{-\mu}  \pa_{\mu}  \phi(x)  |  \leq &  h  L^{-d(j+1)/2} L^{-(j+1)} \|   \phi\|_{\Phi_{j+1}(U^*)}\\
\end{split}
\end{equation}
The factor   $ L^{-d(j+1)}$  compensates  the sum over $x \in U$  and taking   $   L^{-(j+1)} \leq  L^{-1}$
 one obtains for the strong norm and hence the weak norm
\begin{equation}
\|  \cL_4K(U)  \|_{j+1}  \leq   \cO( L^{-1} )  h^2  |\al|
\end{equation}
However  $|\al|   \leq  \cO(1)  h^{-2}A^{-1}  \|K \|_j$  by lemma \ref{albeta} 
which yields   $\|  \cL_4K  \|_{j+1}   \leq   \cO(L^{-1}) \|K \|_j$.  
\bigskip

\begin{lem}   Let  $L$ be sufficiently large.   Then 
the  operator  $\De  $  is a  contraction with arbitrarily small norm.
 \end{lem}
\bigskip

\pr
By    lemma     \ref{donkey}  
\begin{equation}
\| \De K (U) \|_{j+1}  = \|V(0,  \tilde  \al,  U)\|_{j+1}   \leq   h^2  \| \tilde \al \|_{j+1} =h^2 \sup_{U \in \cB_{j+1}} L^{-(j+1)d}    \| \tilde \al \|_{1,U} 
\end{equation}
But    $\tilde \al(x)=0$  if  $x \in B$  and  $B^*$ is away from the boundary.  Hence it    vanishes  if   $d(x, \pa \La_N) >  2^dL^j$   and so
\begin{equation}
 \| \tilde \al \|_{1,U}  \leq  \cO(1) |\al|  \Big| \{ x \in  U:   d(x, \pa \La_N)
  \leq    2^dL^j\}\Big|  \leq  \cO(L^{(j+1)(d-1) }L^j) |\al|
 \end{equation}
Combining these  with    $|\al|   \leq  \cO(1)  h^{-2}A^{-1}  \|K \|_j$  we  obtain   
$\|  \De   K(U)  \|_{j+1}  \leq   \cO( L^{-1} ) A^{-1}   \|K \|_j$
and  hence  $\|  \De   K \|_{j+1}  \leq   \cO( L^{-1} )   \|K \|_j$.

 \subsection{  }

We  now choose      $\tilde  E(B), \tilde   \si$   so  the $V$ terms in (\ref{grrr}) cancel.  First note that
\begin{equation}
\begin{split}
 V^\#(E, \si, B, \phi)   
 = &E(B) + \int   \frac {\si}{4}  \sum_{x \in B}    \sum_{\mu}  (\pa_{\mu}\phi(x) + \pa_{\mu}  \zeta(x)) ^2
   d \mu_{\Ga_{j+1}} (\zeta)\\
= & E(B) +     \frac{\si}{4}  \sum_{x \in B}  \sum_{\mu}       \pa_{\mu}\phi(x)  ^2  
     +     \frac{\si}{4}    \sum_{x \in B}  \sum_{\mu}  ( \pa_{\mu}  \Ga_{j+1}\pa_{\mu}^*)(x,x)\\
   \equiv  & V(E, \si, B, \phi)   +   \frac{\si}{4}     \sum_{\mu}   Tr ( 1_B   \pa_{\mu} \Ga_{j+1} \pa^*_{\mu})  \\
\end{split}
\end{equation}
Thus  the constant terms cancel   if we define $\tilde E  = \tilde E(E, \si, K)$ by 
\begin{equation} 
 \tilde    E(B)   =  E(B)     +  \frac{\si}{4}     \sum_{\mu}   Tr   ( 1_B   \pa_{\mu} \Ga_{j+1} \pa^*_{\mu})  
 +   \beta(K,B)     
\end{equation} 
The   second  order    terms vanish  if we  define  $\tilde \si = \tilde \si (\si, K)$ by
\begin{equation}   
  \tilde   \si =      \si     +   \al  (K)
\end{equation}
Note that we are canceling the constant term  exactly for all $B$,   but  for 
the quadratic term  we  are only canceling exactly  the  invariant  version  away from the boundary.

By  composing   $K' =  K'( \tilde E, \tilde   \si, E, \si, K)$  with  $\tilde E =\tilde  E(E, \si,  K)$  
  and   $\tilde \si   = \tilde  \si(\si ,  K)$ 
we obtain  a new map    $K' = K'(E, \si, K)$.      We  also have new quantities  
$E'(E, \si,  K)$  defined by  
 $E'(B' )  =  \sum_{B \subset B'}\tilde  E(B)$  and  $\si'  =  \si'(\si, K)$   defined by  $\si'=  \tilde \si   =   \si     +    \al(K) $.  These quantities satisfy   (c.f.  (\ref{RG}))
\begin{equation}  \label{noodle}
 \mu_{\Ga_{j+1}} * \left(  I(E,\si)     \circ  K  \right)(\La) =  \left( I'(E',\si')     \circ  K'  \right)(\La) 
 \end{equation}

  We  continue to assume that  $L$ is sufficiently large, and that  $A$ is sufficiently large depending on $L$.

  \begin{thm}  {  \  }
 \begin{enumerate}
 \item For        $R>0$    there   is a $ r>0$  such that   the following holds  for all j.  
  If    $  \|E\|_j,  | \si|,  \| K \|_j  < r$
   then   $ \|E'\|_{j+1}, |   \si' |,   \| K'\|_{j+1}  <  R$.  Furthermore   $E', K', \si' $   are   smooth functions  of   $ E, \si, K$    on this domain with derivatives bounded uniformly in $j$.
 \item   
 The      linearization   of  $K' = K'(E, \si, K)$  at the origin is the   contraction  $\cL K$ where
 \begin{equation}
 \cL=  \cL_1  + \cL_2 + \cL'_3 +  \cL_4  + \De
\end{equation}
\end{enumerate}
\end{thm}
\bigskip   

\pr    For the first  part it suffices to show that the linear maps  $\tilde E =\tilde  E(E, \si,  K)$  
and   $\tilde \si   = \tilde  \si(\si ,  K)$    have norms  bounded uniformly  in   $j$. 
The bound on  $\tilde \si$ follows from   $|\al(K)|   \leq  \cO(1)  h^{-2}A^{-1}  \|K \|_j$ from  lemma \ref{albeta}. 
 The  bound   
on  $\tilde E$ follows  from  the bound on $\| \beta(K)\|_j  \leq  \cO(1) A^{-1}  \|K \|_j$
from  lemma \ref{albeta}   and  the estimate  (\ref{imp})  which gives  for   $B \in \cB_j$
\begin{equation}
\left|  \frac{\si}{4}     \sum_{\mu}   Tr   ( 1_B   (\pa_{\mu} \Ga_j \pa^*_{\mu})  \right|
\leq \cO(1)   |\si|   \sum_{x \in B}  L^{-dj}  \leq   \cO(1)   |\si|  
\end{equation}
Together they  imply    that  $\tilde E  = \tilde E(E, \si, K)$ satisfies
\begin{equation}  \label{snoring}
\| \tilde  E \|_j   \leq   \|E\|_j  +    \cO(1) ( |\si|   +  A^{-1}  \|K \|_j  )
\end{equation}

The  second part  follows  since the linearization  of the new function $K'$  is the linearization 
of the old function  $K'$ composed  with $\tilde E =\tilde  E(E, \si,  K),   \tilde \si   = \tilde  \si(\si ,  K)$.
(All vanish at zero.)   This effects the cancellation and leaves us with  $\cL K$.

 \subsection{  }
 It is convenient to decouple the energy from the other variables.  
 Suppose we start with $E(B)=0$ in (\ref{noodle}).  Then  
\begin{equation}  \label{nokiss}
 \mu_{\Ga_{j+1}} * \left( I(0,\si)   \circ  K  \right)(\La_N) 
 =  \left(  I'( E',\si')  \circ  K' \right)(\La) \\
 \end{equation}
where            $\si'  = \si'( \si, K) $  and  $K' = K'(0, \si, K)$
and    $E'  =E'(0, \si, K) $.
Next remove the  $E'$  making an adjustment  in $K'$.  We   relabel everything with a  plus  and   write      
\begin{equation}  \label{kiss}
 \mu_{\Ga_{j+1}} * \left( I(0,\si)   \circ  K  \right)(\La_N) 
 = \exp \left(   \sum_{B'  \in \cB_{j+1}(\La_N)}  E^+(B') \right) \left(  I'(0,\si^+)     \circ  K^+ \right)(\La_N) 
 \end{equation}
where  
\begin{equation}   \label{a}
\begin{split}   
E^+(\si, K, B') \equiv &  E'(0, \si, K, B')  =  \sum_{B \subset  B'}  \tilde  E( 0, \si, K, B)   \hs    B'  \in  \cB_{j+1}  \\
    \si^+(\si , K)  \equiv &  \si'(\si ,K)  =      \si+   \al(K)     \\
    K^+(\si, K, U)  \equiv    &\exp \left(  - \sum_{ B' \subset  U} E^+(B')   \right) K'(0,\si, K,  U)  \hs   U \in \cP_{j+1}\\
\end{split}
\end{equation}
The dynamical  variables are  now     $ \si^+(\si , K)$    and   $ K^+(\si,K)$.  
 The   energy       $ E^+(\si,  K)$  is driven by the other variables. 
 Since  everything vanishes at the origin    the linearization of  $K^+(\si, K)$  is  still  $\cL K$.  The bound  (\ref{snoring})
 on $\tilde E$   gives  a bound  on  $E^+$
and   our  main theorem becomes:
  
 \newpage 
  \begin{thm}  { \ }
 \begin{enumerate}
 \item For       $R>0$    there   is a $r >0$  such that the following holds  for all j.  
  If     $| \si|,  \| K \|_j  <   r$
   then   $|\si^+ |,   \| K^+\|_{j+1}  <  R$.  Furthermore   $ \si^+, K^+ $   are   smooth functions 
    of   $ \si, K$    on this domain with derivatives bounded uniformly in $j$.
 \item   The extracted energies satisfy 
 \begin{equation}  \label{singsong}
\| E^+( \si, K)\|_{j+1}  \leq   \cO(L^d) \Big(  |\si|   + A^{-1}  \|  K \|_j  \Big)
 \end{equation}
 \item   
 The      linearization   of  $K^+$  at the origin is the contraction   $\cL$.
\end{enumerate}
\end{thm}

\section{The  stable manifold}  \label{flow}

Now  we  etablish  the existence of a stable manifold for the flow.   
 For now we   do not   specialize to the dipole  gas,  but  take  a  
general  initial    point   $\si_0,  K_0$   corresponding to 
an    integral     $\int  (I(0, \si_0)\circ  K_0) (\La_N)  d \mu_{C_0}$.

We assume   $K_0(X, \phi)$ has the lattice symmetries      and satisfies   the conditions  (\ref{special}).
We  also  assume   $|\si_0|,   \| K_0\|_0  < r $  where   $r$   is small enough so the last
theorem holds,  say with $R =1$,  and we can take the first step.
We  apply the transformation (\ref{kiss}) for  $j=0,1,2, \dots$   and continue as long as we can.
This     generates a sequence 
  $\si_j, K^N_j(X)$    by  $\si_{j+1}  =  \si^+(\si_j,  K^N_j)$    and   $K^N_{j+1}  = K^+ (  \si_j, K^N_j)$
 with    extracted energies      $ E^N_{j+1}   =    E^+(\si_j,  K^N_j) $.
Then  we  have with  $I_j(\si_j)  = I_j(0, \si_j)$ for any $k$
 \begin{equation}  \label{ksteps}
\int   (I_0(\si_0) \circ  K_0 )(\La_N)  d \mu_{C_0}
   =    \exp \left(  \sum_{j=1}^k \sum_{B \in \cB_j(\La_N)}     E^N_j(B)   \right)
 \int   (I_k(\si_k) \circ  K^N_k )(\La_N) d \mu_{C_k}
\end{equation}

 The quantities      $K^N_j(X)$   and  $E^N_j(B)$  are independent of  $N$ and 
  have  the lattice symmetries   if   $X,B$  are away from   $\pa  \La_N$ in the sense that
  they have no boundary  blocks.     These properties  are true initially  and  are preserved by the 
  iteration.    In this case  we denote these  quantities by  just    $K_j(X)$   and  $E_j(B)$

By  our  construction  $\al$ defined  in (\ref{sunny}),(\ref{verysunny})  only depends on   $K_j$   and  splitting  $K^+$  into a  linear  
and  a higher order piece  the sequence   $\si_j, K^N_j(X)$    is generated by   the RG transformation
\begin{equation}  \label{flow1}
\begin{split}
\si_{j+1} = &  \si_j + \al(K_j)  \\
K^N_{j+1}  =&  \cL(K^N_j)   +  f(\si_j, K^N_j)\\
\end{split}
\end{equation}
This is regarded as a mapping from the Banach space  $\bbR  \times  \cK_j(\La_N)$
to the Banach space    $\bbR  \times  \cK_{j+1}(\La_N)$
The function
 $  f = f_j$  is smooth  with  derivatives bounded  uniformly   in  $j$
 and satisfies   $f(0,0) =0$,  $Df(0,0)=0$.
 
 For this mapping we can use  the stable manifold theorem  proved  in Brydges  \cite{Bry07}
 to  obtain:

\begin{thm}   \label{stable}  Let  $L$ be sufficiently large,  $A$ sufficiently large  (depending on $L$),
and  $r$ sufficiently small (depending on $L,A$).
 Then there is     $0 < \rho <  r$  and     a smooth  real-valued   function  
     $  \si_0  = h(K_0),\  h(0) =0$, 
 mapping          $\|  K_0 \|_0  < \rho$   into  $|\si_0| < r$  such that 
with these start values   the sequence
$\si_j, K^N_j$    is   defined for all  $0 \leq j \leq  N$  and    
\begin{equation} \label{kings}
|\si_j|   \leq  r 2^{-j}   \hs   \|K^N_j \|_j  \leq     r 2^{-j} 
\end{equation}
Furthermore  the  extracted    energies satisfy  
\begin{equation}  \label{energy}
\|  E^N_{j+1}  \|_{j+1}  \leq    \cO( L^d)   r 2^{-j} 
\end{equation}
\end{thm}
\bigskip

\noindent 
\pr     We  first  establish   the theorem for the invariant  quantities    $K_j(X), E_j(B)$
away from  the boundary.     In this  case    the  RG  transformation  (\ref{flow1})  can be regarded    as   a map 
 from the Banach space  $\bbR  \times  \cK_j(\bbZ^d)$
to the Banach space    $\bbR  \times  \cK_{j+1}(\bbZ^d)$,  since  any  $X  \in \cP_{j,c}(\bbZ^d)$  is 
well  inside  $\La_N$  for $N$ sufficiently large.
 On this space  the   transformation  can       be iterated 
indefinitely.     Furthermore        $\cL$ has the form  $\cL  = \cL'  + \De$   where  $\cL' =  \cL_1 + \cL_2 +  \cL'_3 + \cL_4$
and  where    $\De$ vanishes away
from the boundary.   Thus   the RG  transformation  on the invariant quantities is   
\begin{equation}  \label{flow2}
\begin{split}
\si_{j+1} = &  \si_j + \al(K_j)  \\
K_{j+1}  =&  \cL'(K_j)   +  f(\si_j, K_j)\\
\end{split}
\end{equation}
Both     $\cL', \al$ are   contractions  with arbitrarily small norm  for  $A,L$ large.
Then   we can apply the stable manifold  theorem from     \cite{Bry07}, Theorem 2.16  with  parameters  $\mu = 1/2$  and $\al=1$.
This  yields  the function  $\si_0  = h(K_0)$ and with these initial values the  sequence 
$\si_j, K_j$  satisfies  (\ref{flow2})   with the bounds  (\ref{kings}).

Once  we know that  $\si_j$  is not growing  we  can give a direct  proof  that
$\|  K^N_j  \|_j$   satisfies the bound (\ref{kings})  reproducing the   results for  $K_j$ 
but  now including the  boundary polymers.
The bound  is true initially  since  $K^N_0(X)  = K_0(X)$
even  if  $X$  touches the boundary. Suppose it is true  for   $j$.   We  have  
$K^N_{j+1}  =  \cL(K^N_j)   +  f(\si_j, K^N_j)$    where  $\cL$  is a contraction with  norm
less  than  $1 /4$   and   $ f(\si_j, K^N_j)$  is second order.    
Hence  for some constant   $M$   and    $r$  sufficiently small
\begin{equation}
\begin{split}
\|  K^N_{j+1}  \|_{j+1}     \leq   &\ \frac14 \|  K^N_j \|_j  +  M \Big(   |\si_j|^2  +  \|K^N_j \|_j^2  \Big) \\
\leq  &\ \frac14   \Big(     r 2^{-j}  \Big)
+   2M   \Big(     r 2^{-j}    \Big)^2\\
\leq  & \     r 2^{-j-1} 
   \\
\end{split}
\end{equation}
which is the bound for $j+1$.

Finally the  energy     bound  (\ref{energy})
comes from   the bounds  on  $\si_j,  K^N_j$   and    (\ref{singsong}).

\section{The dipole gas}    \label{result}
\subsection{the initial  density} 

We  now specialize to the dipole gas  and complete the proof of the theorem. 
The first issue is to  adjust the dipole gas density so it  becomes a point on the stable manifold.

For the dipole gas the initial density  $\cZ^N_0 = \cZ^N_0(z,\si)$ is given in  (\ref{picky}).
We   break it into pieces  defining    for  $B \in \cB_0$,      $W_0(B)  =  z W(\sqrt{1+ \si_0},  B)$  as  in    (\ref{syrup}) 
and  $V_0(B)  =  \si_0V( B)$  as  in      (\ref{64}).   Then we follow with   a Mayer expansion 
to   put the density in the form we want.
\begin{equation}  \label{sink}
\begin{split} 
 \cZ^N_0   =&    \prod_{B \subset  \La_N} e^{  W_0(B)-V_0(B)} 
 =    \prod_{B \subset  \La_N} \Big(e^{ - V_0(B) } + (e^{  W_0(B)}-1)e^{ - V_0(B)}  \Big) \\
=&   \sum_{X \subset   \La_N } I_0(\si_0,\La_N - X)  K_0(X) 
=  (   I_0(\si_0) \circ  K_0 )(\La_N)\\
\end{split}
\end{equation}
where  $ I_0(\si_0,B)=  e^{ - V_0(B)}$   and   $K_0(X)   = K_0( z,  \si_0, X)$   is  given by  
\begin{equation}
K_0( X )  = \prod_{B  \subset  X}  (e^{  W_0(B)}-1)e^{- V_0( B)}
\end{equation}
Note that     $K_0$ has the lattice symmetries      and satisfies   the conditions  (\ref{special}).  To start 
the flow we need:

\begin{lem}  \label{stunted}
Given  $r>0$  if  
if    $|z|$  and  $ |\si_0|,   $  are sufficiently small  then   $\|  K_0(z,  \si_0)  \|_0  \leq  r$.
Furthermore   $K_0$ is a smooth function of  $(z,  \si_0)$.
\end{lem}
\bigskip

\pr   Consider  the $G=1$   norm   $\| \cdot  \|_{00}$ defined in  (\ref{mmm}).   As in    
 (\ref{threethree})  we have  
\begin{equation}       \label{taffy}  
\|  e^{W_0( B)} -1 \|_{00}  \leq  
   \exp  \Big( 2  |z|  e^{h  \sqrt{d(1+ \si_0)}}  \Big)  -1   \leq    c |z|
\end{equation}
for some  constant  $c$.
Also   by  lemma  \ref{donkey}    $\|  e^{-V_0(B ) }\|_{s,0}        \leq   2$.
Combining these 
\begin{equation}
\| ( e^{W_0(B)}-1) e^{-V_0(B )} \|_{s,0}  
    \leq  
\|  e^{W_0(B)}-1\|_{00}   \|  e^{-V_0(B )} \|_{s, 0}    
\leq   2c|z|
\end{equation}
Then  
\begin{equation}
\|K_0(X) \|_{s,0}   \leq   \prod_{B  \subset  X}   \| ( e^{W_0(B)}-1)e^{-V_0(B)} \|_{s,0}
\leq   ( 2c|z| )^{|X|_0}  
\end{equation}
Then  same follows   for the weak   norm $\|K_0(X) \|_0$
and so  
\begin{equation}
\| K_0 \|_0  =  \sup_{X \in \cP_{0,c}}   \|K_0(X) \|_0  A^{|X|_0}  \leq    \sup_{X}  (2c|z|A)^{|X|_0}   \leq 2c|z|A    <    r
\end{equation}

The smoothness follows similarly from lemma   \ref{donkey}  and  lemma  \ref{burro}.   For example   consider
 the part of    $K_0$  depending on  $W$  which is
\begin{equation}
K'_0(X)  = \prod_{B  \subset  X}  (e^{   W_0(B)}-1)
\end{equation}
We show that   the derivative with respect to  $\si_0$   has a finite norm.   The derivative is computed 
as  
\begin{equation}
\frac{\pa  K'_0(X)} {\pa \si_0}  =   \sum_{B_0 \subset X}   z W'(  \sqrt{ 1 + \si_0},  B) \frac{1}{2\sqrt{1+ \si_0}}
\prod_{B \subset  X-B_0}   (  e^{   W_0(B)}-1)
\end{equation}
Then  by   (\ref{www})   and    (\ref{taffy})    we have for some constant  $c'$
\begin{equation}  \label{order}
\|\frac{\pa  K'_0(X)} {\pa \si_0}\|_{s,0}   \leq     \sum_{B_0 \subset X}  | z| \| W'(  \sqrt{ 1 + \si_0},  B)\|_{s,0} 
\prod_{B \subset  X-B_0}    \| e^{   W_0(B)}-1\|_{00}       \leq   (  c' | z|  )^{|X|_0} 
\end{equation} 
and so   
\begin{equation}
\|\frac{\pa  K'_0} {\pa \si_0}\|_0  \leq   Ac'|z| 
\end{equation}
The other pieces may be treated similarly.   \footnote { For  $  \pa K_0/ \pa  \si_0$  we must   combine the  estimate  (\ref{order})  with    estimates
 $\|  e^{-V_0(B ) }\|_{s,0}        \leq   2$.  For this  use   $G_{s,0}^2  \leq   G_0$. }
 This completes the  proof.
\bigskip

To  apply   theorem   \ref{stable}    we         need  to choose $ \si_0$   so that
 $\si_0  =   h(  K_0(z, \si_0))$.   
 
 \begin{lem}
 The equation   $\si  =   h(  K_0(z, \si ))$  defines   a  smooth  implicit function  $\si  = \si (z)$
 near the origin which satisfies  $\si(0) =0$.
 \end{lem}
 \bigskip
 
 \pr  Let      $f(z, \si)  =  \si-   h(  K_0(z, \si))$.  Then  $f(0,0) =0$.  
 The function  $h$ is smooth by theorem \ref{stable}  and the function $K_0$ is 
 smooth by  lemma  \ref{stunted}.   Hence  $f$ is smooth  and we compute
 \begin{equation}
 f_{\si}(0,0)   =   1 - Dh(  0; ( K_0)_\si(0,0))
 \end{equation}
  But    $K_0(0, \si )  =0$,  hence   $ (K_0)_\si(0,0)  =0$  and hence  $ f_\si  (0,0)   =  1   \neq 0$.
 By the   implicit function theorem  there exists  $\si = \si(z)$  so that  $f(z, \si(z)) =0$.
 This completes the proof.
 \bigskip

 Taking  $|z|$  sufficiently small and 
making the   choice    $\si_0 = \si(z)$    the start density  $I_0( \si(z))  \circ  K_0(z, \si(z) )$   is now tuned and we  can apply  theorem \ref{stable}.    
We  have  for  $0 \leq  k \leq  N$
 \begin{equation}    \label{sinking}
 \begin{split}
 Z'_N(z,  \si(z)  )=& \int \Big(I_0( \si(z))  \circ  K_0(z, \si(z)) \Big)(\La_N) d \mu_{C_0}\\
   = &   \exp \left(  \sum_{j=1}^k \sum_{B \in \cB_j(\La_N)}     E^N_j( B)   \right)
\int   (I_k(\si_k) \circ  K^N_k )(\La_N) d   \mu_{C_k}\\
\end{split}  
\end{equation}
where   $|\si_j|  \leq  r2^{-j}$  and  $ \| K^N_j\|_j  \leq  r2^{-j}$  
and    $\|E^N_{j+1}\|_{j+1}  \leq  \cO(L^d)  r 2^{-j}$.

\subsection{the pressure}

Now we  can  show  the pressure  has an infinite volume limit,
completing    the proof of theorem  \ref{one}.

\begin{thm}  \label{six}
For    $|z|$  sufficiently small    the following  limit  exists:
\begin{equation}
  \lim_{N\to \infty}  |\La_N|^{-1}  \log  Z _N'(z,  \si(z))   
  \end{equation} 
 \end{thm}
 \bigskip

\pr 
Take      $k=N$  in   (\ref{sinking}).    At this level there is only one     block  $\La_N \in \cB_N(\La_N)$ and so   
\begin{equation}  \label{still}
\begin{split}
  |\La_N|^{-1}  \log  Z '_N(z, \si(z))   =&   |\La_N|^{-1}  \sum_{j=1}^{N} \sum_{B \in \cB_j(\La_N)}     E^N_j(B)   \\
     + &   | \La_N|^{-1} \log \left(   \int \left[  I_N(\si_N, \La_N)   +   K^N_N (\La_N) \right]d \mu_{ C_N}  \right)\\
\end{split}     
\end{equation}

The second  term has the form 
\begin{equation}  \label{ugh}
  | \La_N|^{-1} \log  \left(  1   +  \int       F_N d \mu_{ C_N}  \right)
\end{equation}
where
\begin{equation}
F(\La_N) =    I_N(\si_N, \La_N)  -1  +    K^N_N(\La_N)
\end{equation}
By   (\ref{I})    and  (\ref{oliver}) 
\begin{equation}
\|  I_N(\si_N, \La_N)  -1 \|_N   \leq  \cO(1)h^2 |\si_N|    \leq    \cO(1)h^2   r2^{-N} 
\end{equation}
 and  
\begin{equation}
\|K^N_N(\La_N)\|_N   \leq A^{-1}  \|K^N_N \|_N  \leq  A^{-1}r 2^{-N}
\end{equation} 
so that  $\| F(\La_N) \|_N$   is   $\cO(2^{-N})$  as  $N \to \infty$.

In a following lemma we prove that  for  $h$  sufficiently large 
$\int   G_N(\La_N,0, \zeta)  d \mu_{C_N} (\zeta)  \leq     2 $.
Then we estimate
\begin{equation}
\left|\int        F_N(\La_N ) d \mu_{ C_N}\right |  \leq        \|  F(\La_N) \|_N \int   G_N(\La_N,0, \zeta)
  d \mu_{C_N} (\zeta)  \leq  2  \|  F(\La_N) \|_N
=    \cO(2^{-N}) 
\end{equation}
Hence   the expression  (\ref{ugh})  is  $  \cO(2^{-N})   | \La_N|^{-1} $
 and  goes to zero very  quickly  as  $N  \to  \infty$ 
 
Now we consider the first term   in (\ref{still}).  If we replace 
$E^N_j(B)$  by the invariant  quantity  $E_j(B)$  we have  
\begin{equation}
 |\La_N|^{-1}  \sum_{j=1}^{N} \sum_{B \in \cB_j(\La_N)}     E_j(B)
  = L^{-dN} \sum_{j=1}^{N }  L^{d(N-j)}E_j(B)  =  
   \sum_{j=1}^{N }  L^{-dj}E_j(B) 
 \end{equation}
  Since  $|E_j(B)|   =  \cO(2^{-j})$  this converges to the infinite sum  as  $N \to \infty$.

 Now we are left with 
\begin{equation}
 |\La_N|^{-1}  \sum_{j=1}^{N} \sum_{B \in \cB_j(\La_N)}   (  E^N_j(B) -E_j(B))
\end{equation} 
 Since  $ E^N_j(B) -E_j(B)$ vanishes away from the boundary
the  term is bounded by   a constant times
\begin{equation}
\begin{split}
& |\La_N|^{-1}  \sum_{j=1}^{N}  \sum_{B \in \cB_j(  \pa  \La_{N})} 2^{-j} 
   \leq  \cO(1)   L^{-dN}  \sum_{j=1}^{N}    L^{(d-1)(N-j) } 2^{-j} \\
&\leq  \cO(1)  L^{-N}\sum_{j=1}^{N} L^{-(d-1)j}  2^{-j}
=     \cO(  L^{-N})\\
\end{split}
\end{equation}
where   $ \cB_j(  \pa  \La_{N})$ are the boundary blocks in  $ \cB_j(    \La_{N})$.
Hence   this  goes to zero as  $N \to \infty$  to complete the proof,  except for the next lemma.

\begin{lem}   For  $h$ sufficiently large
\begin{equation}  \label{missing} \int   G_N(\La_N,0, \zeta)  d \mu_{C_N} (\zeta)      \leq   2
\end{equation}
\end{lem}
\bigskip

\re    The proof is similar to the 
bound on   $\int  G_j(X, 0, \zeta)  d\mu_{\Ga_j}$  given  in    \cite{Bry07}, (6.53).   However    
\begin{equation}
C_N(x-y)  =   \sum_{j=N+1}^{\infty}  \Ga_j(x-y)
\end{equation}  
has  infinite range and so we must approach things a little differently.  
Note that   $C_N$ does satisfy essentially the same bound as   $\Ga_{N+1}$ namely 
\begin{equation}  \label{comely}
|\pa^{\al}C_N(x)  |
\leq  2 c_{\al} L^{-(d-2 + |\al|)N}
\end{equation}
\bigskip

\pr 
 As  noted  in   \cite{Bry07}, lemma 6.31,  
after a Sobolev inequality and a Holder inequality it suffices to 
show that for fixed  $a$  and   any multi-index  $\al$  that  
\begin{equation}
\int  \exp \left (   ah^{-2}   L^{(2|\al| -2)N}  \sum_{x  \in  \La^*_N}  |(\pa^{\al}\zeta)(x)|^2 \right) d \mu_{C_N}(\zeta)
\leq    \exp  ( \cO(h^{-2} )
\end{equation}
With   
\begin{equation}
A  =  2 ah^{-2}   L^{(2|\al| -2)N}\   C_N^{1/2} (\pa^{\al})^*  1_{ \La^*_N} \pa^{\al}C_N^{1/2}
\end{equation}
The integral is computed as     
\begin{equation}
\int  \exp\left( \frac12  (\zeta, A  \zeta) \right) d \mu_{I}(\zeta)
=   \det  (  1  +  A)^{-1/2}
\end{equation}
provided  $A$ is trace class.   But  by   (\ref{comely})  and  $|\La_N^* |  \leq  \cO(1) | \La_N|$  we 
have   for some constant  $k$
\begin{equation}
\begin{split}
\tr  (A)   = & 2 ah^{-2}   L^{(2|\al| -2)N}\ \tr (     1_{ \La^*_N} \pa^{\al} C_N (\pa^{\al})^*)   \\
= &2   ah^{-2}   L^{(2|\al| -2)N} \sum_{x \subset  \La^*_N}   (-1)^{|\al|}  (\pa^{2\al }  C_N)(0) \\
\leq   &  4 ah^{-2}  c_{2\al} \sum_{x \subset  \La^*_N}    L^{-dN}  
\leq    kh^{-2} \\
\end{split}
\end{equation}
Then  also  $\|A\|  \leq  kh^{-2}$  and  so $\tr  (A^n)  \leq   \| A \|_1  \|A\|^{n-1}   \leq  k^nh^{-2n}  $.
Now as  in  (\ref{sonny})  we have   
\begin{equation}
 \det  (  1  +  A)^{-1/2}  = \exp \left (   \frac12  \sum_{n=1}^{\infty}  \frac{(-1)^{n}}{n}  \tr(  A^n) \right)   
  \leq       \exp \left (     \sum_{n=1}^{\infty}    k^nh^{-2n}  \right)    \leq     \exp  (   \cO(h^{-2}))   
\end{equation}
This completes the proof.
 \bigskip

\appendix

\section{Degenerate Gaussian measures}  \label{App}
 In the text  we  use  degenerate   Gaussian measures.   Here we give a precise definition.
 
 Let    $\Ga$   be  a bounded symmetric  operator  on  real-valued $\ell^2(\bbZ^d)$  that  is 
 positive in the sense  that  
 \begin{equation}
 (f,\Ga f)   =  \sum_{x,  y  \in \bbZ^d} f(x)  \Ga(x,y)   f(y)   \geq   0
 \end{equation}
 but only semi-definite because we   allow the possibility that   $(f, \Ga  f)  =0$  for some  $f \neq  0$.

We  want to consider a   Gaussian  process   with  covariance  $\Ga$. Since    it  is  only  semi-definite   this is not quite standard.   A convenient way to 
proceed  is  to   let   $Z(x)$ be  a Gaussian
process indexed  by  $x  \in \bbZ^d$    with identity covariance,  i.e.    $Z(x)$ are   independent normal random variables. 
Let   $(\cM,  \mu)$ be the   underlying   measure space.  Let   $\Ga^{1/2}(x,y)  =  (\de_x,  \Ga^{1/2}  \de_y)$  be the kernel  of  $\Ga^{1/2}$ 
and define   $\phi  =  \Ga^{1/2}  Z$ by  
\begin{equation}
\phi(x)   =  \sum_y   \Ga^{1/2}(x,y)  Z(y)
\end{equation}
This  sum converges in the $L^2(\cM, \mu)$  since 
\begin{equation}
\sum_y   | \Ga^{1/2}(x,y)|^2   =   \Ga(x,x)   < \infty
\end{equation}  
Expectations  are   integrals  
$\int     [ \cdots  ] d \mu$     and   we   use the notation 
\begin{equation}
\int   F(  \phi) \  d \mu_{\Ga} (\phi)   \equiv    \int   F( \Ga^{1/2}Z) \  d \mu(Z)
\end{equation}
when the integral  exists.  
In  particular  if   $\phi(f)  =  \sum_x  \phi(x)f(x)$ with  $f  \in \ell^2(\bbZ^d)$ we have the 
characteristic function   
\begin{equation}
\begin{split}
\int   \exp (   i \phi(f)  )     \  d \mu_{\Ga} (\phi) = &   \int     \exp (   i Z(  \Ga ^{1/2}  f)  )        d \mu(Z)\\
=  &      \exp \Big(  - \frac12    \|  \Ga ^{1/2}  f  \| ^2   \Big)     \\
=  &      \exp \Big (  - \frac12   (f , \Ga  f  )  \Big )     \\
\end{split}
\end{equation}
  which  verifies that  $\phi$ is     a Gaussian  process  with covariance  $\Ga$.

If  $\phi_1= \Ga_1^{1/2}Z_1 $  is  Gaussian with  covariance  $\Ga_1$ on $(\cM_1, \mu_1)$
 and  $\phi_2= \Ga_2^{1/2}Z_2$  is Gaussian with 
covariance  $\Ga_2$  on $(\cM_2, \mu_2)$ ,   then   $\phi_1 + \phi_2$
  on the    product space   $(\cM_1 \times   \cM_2,   \mu_1 \times  \mu_2)$      gives   a  realization 
of a Gaussian    process with covariance   $\Ga   =  \Ga_1 + \Ga_2$.
This works   because the characteristic function is 
\begin{equation}
\begin{split}
&  \int      \exp \Big(   i ( \phi_1(f)  + \phi_2(f) ) \Big)   d \mu_{\Ga_1} (\phi_1)  d \mu_{\Ga_2} (\phi_2)\\
= &      \exp  \Big(  - \frac12   (f , \Ga_1  f  )  \Big)   \exp \Big(  - \frac12   (f , \Ga_2  f  )   \Big)  \\
=&    \exp  \Big(  - \frac12   (f , \Ga  f  )   \Big) \\
\end{split}
\end{equation}

\end{document}